\documentclass[12 pt]{article}
\usepackage{mathrsfs}
\usepackage{jheppub}
\usepackage[table]{xcolor}
\usepackage{mathtools,leftindex,tensor,mhchem}
\usepackage{amssymb}
\usepackage{listings}
\usepackage{xcolor}

\lstdefinelanguage{Wolfram}{
  morekeywords={
N,Range,Abs,Length,Print,Dataset,DeleteDuplicatesBy,KeyTake,
Values,Select,Tuples,Reap,Do,If,With,Module,Row,KeyDrop,Sow,
rhoGravFakeHeavyLabelData,rhoGravFakeParallel
  },
  sensitive=true,
  morecomment=[s]{(*}{*)},
  morestring=[b]"
}

\usepackage{color, colortbl}
\usepackage{dsfont}
\usepackage{amsmath}
\usepackage{bbm}
\usepackage{amsfonts}
\usepackage{booktabs}
\usepackage{orcidlink}
\usepackage{multicol} \usepackage{multirow}
\usepackage{physics}
\usepackage{amsmath}
\usepackage{caption}
\usepackage{subcaption,longtable,stmaryrd}
\usepackage{slashed}
\usepackage{bigints}
\usepackage{cancel}
\usepackage{multicol}
\usepackage{blindtext}
\usepackage{tikz}
\usepackage{enumitem}
\usepackage[customcolors,shade]{hf-tikz}
\usepackage{graphicx}
\usepackage{cancel}

\DeclareMathAlphabet\mathbfcal{OMS}{cmsy}{b}{n}
\DeclareSymbolFont{usualmathcal}{OMS}{cmsy}{m}{n}
\DeclareSymbolFontAlphabet{\mathcal}{usualmathcal}
\DeclareSymbolFont{rmlargesymbols}{OMX}{mdbch}{m}{n}
\DeclareMathSymbol{\rmintop}{\mathop}{rmlargesymbols}{82}
\DeclareMathSymbol{\rmointop}{\mathop}{rmlargesymbols}{72}
\newcommand{\rmint}{\rmintop\nolimits}
\definecolor{mygray}{gray}{0.5}

\title{\boldmath{Pure gravity OPE density for genus-two handlebody in $\text{AdS}_3$}}

\author[a,b]{Saptaswa Ghosh\,\orcidlink{0000-0002-0237-7042}}
\author[b]{and Sounak Pal\,\orcidlink{0000-0002-2250-0466} }
\affiliation[a]{\it Centre for High Energy Physics, Indian Institute of Science,
C.V. Raman Avenue, Bangalore 560012, India.}
\affiliation[b]{\it Indian Institute of Technology, Gandhinagar, Gujarat-382055, India}

\emailAdd{ghoshsaptaswa22@gmail.com}
\emailAdd{palsounak@iitgn.ac.in}
\abstract{In this work, we compute the operator product expansion density for the genus-two handlebody by exploiting its duality with extremal conformal field theories \cite{Yin:2007gv}. Using a Poincaré construction, we analytically determine the first- and second-order corrections in the pinching parameter to the genus-two OPE density through a direct inversion of the partition function. As a consistency check, we also compute the genus-one density within the same Poincaré framework and find perfect agreement with the result obtained from the lightcone bootstrap analysis \cite{Benjamin:2019stq}. It is known that, at genus one, the OPE density of extremal CFTs exhibits a negativity pathology for states lying just above the black-hole threshold. We show that this pathology continues to persist at genus two, despite the emergence of a nontrivial inversion kernel. In particular, the OPE density associated with states of both odd and even spin $j_i$ remains negative within a very narrow band above the black-hole threshold. We further analyze the sharp transition from large positive to large negative values of the OPE density across a curve along which the density vanishes.
}
\makeatletter
\begin{document}
\maketitle
\flushbottom
\section{Introduction}
The problem of formulating a consistent theory of gravity has remained one of the
central challenges in theoretical physics for several decades. While many attempts have been made in certain cases, the construction of tractable and well-defined models of
quantum gravity continues to present notable obstacles. Much of the progress has occurred in lower-dimensional settings, where simplifications enable explicit computations. A particularly important achievement in this regard was the formulation of Jackiw–Teitelboim (JT) gravity in two dimensions \cite{Jackiw:1984je,Teitelboim:1983ux,Maldacena:1997re,Iliesiu:2020qvm,Iliesiu:2019lfc,Moitra:2021uiv,Nayak:2018qej}, which has served as an important testing ground for the connections
between gravity, quantum field theory, and random matrix models. The rich field of two-dimensional gravity has also been studied in aspects of de-Sitter space \cite{Okuyama:2025hsd,Verlinde:2024zrh,Verlinde:2024znh}. Recent studies regarding factorization puzzle has considered the end-of-world brane insertions with correlations between them, in lower dimensions \cite{Wang:2025bcx,Blommaert:2021fob}. Complementary insights have also arisen from the AdS/CFT correspondence, which provides a non-perturbative definition of quantum gravity in terms of conformal field theories. In recent years, there has
been a surge of interest in two-dimensional gravity and its deformations, with important contributions to the computation of partition functions \cite{Maloney:2007ud}, spectral densities, and observables
in both perturbative and non-perturbative regimes. Several approaches have been used
to compute partition functions and spectral form factors (SFF) \cite{Cotler:2016fpe,Okuyama:2023pio} with integrable irrelevant deformations (such as $T\bar T$ deformations\cite{Cavaglia:2016oda,Smirnov:2016lqw}) \cite{Conti:2019dxg,Gross:2019uxi,Brizio:2024doe,Chakraborty:2020xwo,Chakraborty:2021gzh,Bhattacharyya:2023gvg,Ebert:2022gyn,Bhattacharyya:2025gvd}. While SFFs has been calculated for triple-scaled Sachdev-Ye-Kitaev (SYK) models \cite{Kitaev:2015part1, Kitaev:2015part2} (JT gravity) and its cousins, recently complexity and SFF in double-scaled SYK (DSSYK) models has also grown certain interest \cite{Goel:2023svz,Aguilar-Gutierrez:2026jjv,Aguilar-Gutierrez:2025hty}. \\ \par

\noindent
By contrast, the situation in three spacetime dimensions is considerably more intricate. Pure Einstein gravity in three dimensions is non-dynamical: it admits no local propagating degrees of freedom, such as gravitational waves, and hence is dramatically simpler than its higher-dimensional counterparts. This feature suggests that three-dimensional gravity may admit a formulation as a topological quantum field theory (TQFT) \cite{Mikhaylov:2017ngi,Atiyah:1989vu,Witten:2007kt,Collier:2024mgv,Yan:2023rjh}, in which the dynamical content is encoded in edge modes  associated with asymptotic boundaries \cite{Belaey:2025ijg,Blommaert:2018rsf}. Nevertheless, despite this apparent simplification, a complete quantum treatment of three-dimensional gravity remains elusive, in sharp contrast to the large amount of progress achieved in JT gravity. Apart from this, advocating the problem of bulk factorization in 3D gravity, certain edge state calculations for bulk factorization has been done in \cite{Mertens:2022ujr}.\\ \par
A central obstruction arises from the tension between locality and consistency conditions in the holographic dual description. Whereas ensemble averages provide an effective description in JT gravity, in three dimensions, one encounters strong constraints from modular invariance and the operator product expansion. In particular, solving the crossing equations in the holographic large-$c$ regime is a hard problem, and explicit solutions are extremely rare \cite{Benjamin:2019stq}. Over the past few years, notable efforts have been devoted to quantizing three-dimensional gravity which can be described as a topological field theory (TFT) within the framework of the AdS/CFT correspondence \cite{Collier:2023fwi}.\footnote{Certain supersymmetric computations are also performed in super virasoro TQFT \cite{Bhattacharyya:2024vnw,Eberhardt:2026hfh}.} Yet, the precise status of the dual boundary CFT remains a mystery. Motivated by the important lessons of two-dimensional gravity, it has been conjectured that the dual description is not merely a single large-$c$ CFT, but rather an ensemble of chaotic large-$c$ CFTs \cite{deBoer:2024mqg,Chandra:2022bqq,Belin:2023efa,Jafferis:2025jle,Jafferis:2025vyp,Jafferis:2025yxt,Hartman:2025ula,deBoer:2025oge}. Despite many important advances in this direction, no explicit example of a unitary, large-$c$ chaotic CFT with a sparse light spectrum and full Virasoro symmetry has yet been constructed. Different type of ensembles also produce the 3D gravity partition function on the genus-one topolgy \cite{Maloney:2020nni,Afkhami-Jeddi:2020ezh,Benjamin:2021wzr}.
There is a recent surge of interest towards describing the 3D gravity theory using Random-matrix theory (RMT) descriptions \cite{DiUbaldo:2023qli,Boruch:2025ilr}. Recently, some calculations involving ensemble of rational conformal field theories (RCFTs) for reproducing the dual gravity theory in 3D has been done in \cite{Castro:2011zq,Barbar:2023ncl,Dymarsky:2024frx,Dymarsky:2026lnf,Barbar:2025vvf}.  \\ \par
In recent times, significant progress has been achieved by reinterpreting $AdS_3$ quantum gravity in terms of TQFT  \cite{Belin:2026pko,Cotler:2016fpe,Collier:2024mgv,Post:2024itb,Yan:2023rjh,Yin:2007gv,Yan:2025usw, Aghaei:2015bqi,Aghaei:2020otq,Poghosyan:2016kvd,Takahashi:2024ukk}, emphasizing bulk formulations rather than conventional boundary approaches. This perspective opens up a new route to computing observables directly in the bulk, circumventing some of the challenges associated with boundary CFT constructions. It uses irrational conformal field theory (Liouville CFT) techniques \cite{Nakayama:2004vk,Zamolodchikov:2001ah,Ponsot:1999uf,Suchanek:2010kq,Belavin:2007eq}. Further explorations of this line of thought have appeared in \cite{Hung:2024gma}, with related work including the development of simplicial 3D gravity models based on BCFT data \cite{Hung:2024gma,Geng:2025efs}.
\\ \par

The study of two-dimensional conformal field theories (2d CFTs) through their partition functions acquires new depth at genus two. While the torus partition function organises the spectrum and is constrained by modular invariance under $PSL(2,\mathbb{Z})$, the genus-two partition function depends on a higher-dimensional moduli space and is constrained by invariance under the Siegel modular group $Sp(4,\mathbb{Z})$ \cite{Collier:2023fwi}. This enlarged symmetry mixes contributions across different operator channels and probes information beyond the spectrum, tying together conformal dimensions and operator product expansion (OPE) data. A key feature of genus-two surfaces is the presence of distinct degeneration limits. In the \emph{separating degeneration}, the surface splits into two tori connected by a thin tube, and the partition function reduces to a sum over intermediate states weighted by OPE coefficients. In the \emph{non-separating degeneration}, a handle pinches off, corresponding to a sum over operators propagating in a long channel. These limits provide a physical interpretation: light operators dominate the partition function near the boundary of moduli space, while modular invariance determines how this data extends to the interior. The genus-two surfaces has geometric interpretation, possibly in terms of multicentered black holes \cite{Brill:1995jv,Cai:2022qac}. It is useful not to picture several Schwarzschild-like objects sitting at different spatial locations as pure 3D  gravity has no local gravitational propagating degrees of freedom. Hence, all the interesting information is global or topological. Multi-black-hole spacetimes can be constructed as quotients of $AdS_3$ for an appropriate discrete subgroup $\Gamma\in PSL(2,R)$. The resulting geometry can have asymptotic AdS regions (which can be more than one), connected through a common interior. Each exterior can be framed locally as an ordinary BTZ exterior. This is exactly the geometry of the classic multi-black-hole solutions. Therefore, when we calculate the density of such multicentered black hole spacetimes, it constrain the data of the underlying CFT spectrum.\\ \par
In this work, we explore these questions by focusing on the genus-two partition function \cite{Yin:2007gv} and its implications for OPE densities in irrational CFTs and their gravitational duals \cite{Alday:2019vdr}. Our aim is to clarify how modular invariance constrains the extension from light to heavy data and to investigate whether the negative genus-one density around the black-hole threshold persists in the genus-two separating channel. For the genus-one case, negative densities of states have been found for odd spins around the black-hole threshold for pure gravity \cite{Alday:2019vdr,Alday:2020qkm}, although the heavy spectrum shows the universal Cardy scaling. Several proposals have been made to address this issue \cite{Alday:2019vdr,Alday:2020qkm,Bae:2016yna,Benjamin:2020mfz}, including orbifold singularities and Seifert manifolds \cite{Yan:2025usw}. In this work, we compute the handlebody OPE density in the separating degeneration, using direct inversion and state the difference with the obtained density via the Rademacher circle method \cite{LopesCardoso:2021aem,Baccianti:2025gll,Alday:2019vdr}. The explicit sign check is restricted to representative nonzero-spin states in the local $s=1$ approximation and does not prove positivity of the complete genus-two density. Recently, in \cite{Simmons-Duffin:2025qox} genus-two OPE density for a generic CFT has been studied.\\ \par
\noindent
Our paper is organised as follows. In Section~\ref{sec2}, we describe the $g=2$ handlebody partition function for extremal conformal field theory (ECFT). We also describe the seed kernel over which we need to sum its images. In Section~\ref{sec3}, we proceed to perform this sum over the modular images using the Poincaré series. After performing the sum, in Section~\ref{sec4}, we find the OPE density for 2d ECFTs on $g=1$ using direct inversion.  Finally, we comment on the  OPE density for pure gravity for Euclidean saddles at genus-two in Section~\ref{secc5}. Section~\ref{sec5} consists of a conclusion and a few possible future outlooks. Finally, in Appendix~\ref{appA} we derive a generic formula using the Rademacher method for finding the OPE density. In Appendix~\ref{appB} we demonstrate the sum over modular images. We also cast a few specific cases of \textit{Kloosterman sum} in a table.
\section{Genus-two handlebody in separating degeneration and the extremal CFT}\label{sec2}
In this section, we primarily collect all the necessary ingredients to compute the genus-two handlebody density. As a primary simplification, we work near the separating degeneration and retain the block-diagonal modular subgroup relevant in this limit.

\subsection{Brief review of ECFT and partition function at $g=2$}
\begin{figure}[htb!]
    \centering
\includegraphics[width=0.35\linewidth]{{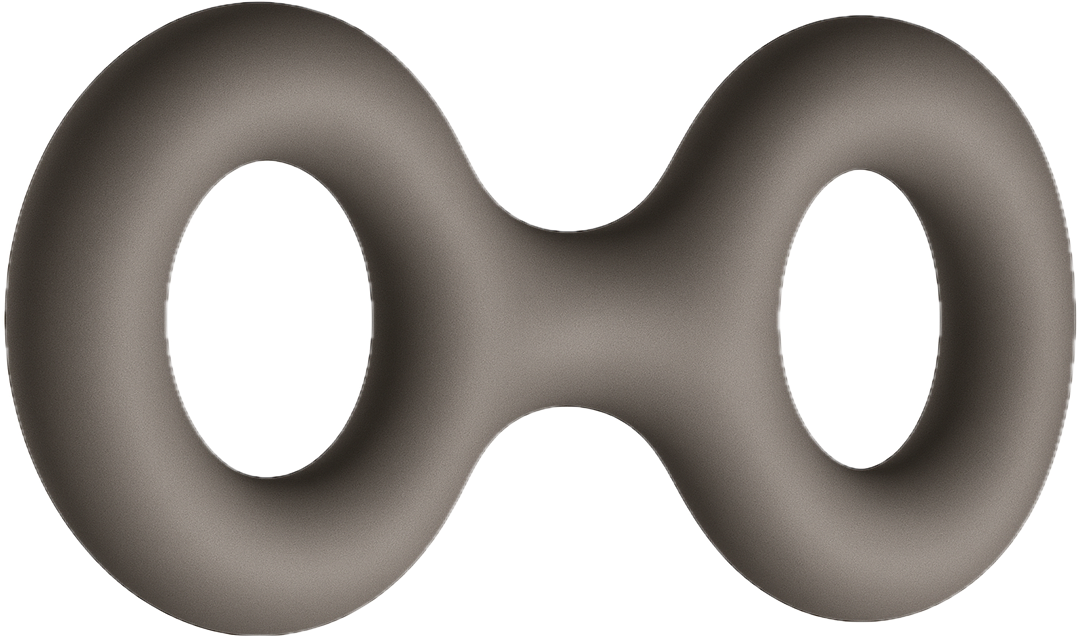}}
    \caption{A genus-two handlebody in pure gravity in $AdS_3$.}
    \label{fig:handlebody}
\end{figure}

\noindent
In this section, we briefly review the genus-two partition function of an ECFT, which comprises of Eisenstein series. The most notable example of ECFT is the Monster CFT \cite{Gaberdiel:2012um,Gaiotto:2008jt,Gaiotto:2007xh} at central charge $c=24$, constructed by Frenkel–Lepowsky–Meurman, whose partition function is the famous \textit{modular $j$-function}. Apart from this case, the exact form of the partition function for extremal CFT(s) at higher central charges remains an open question. Their existence would provide a concrete microscopic definition of pure $AdS_3$ gravity, while their non-existence would place significant and precise constraints on the landscape of consistent theories of quantum gravity.\\

\begin{figure}
    \centering
\includegraphics[width=0.64\linewidth]{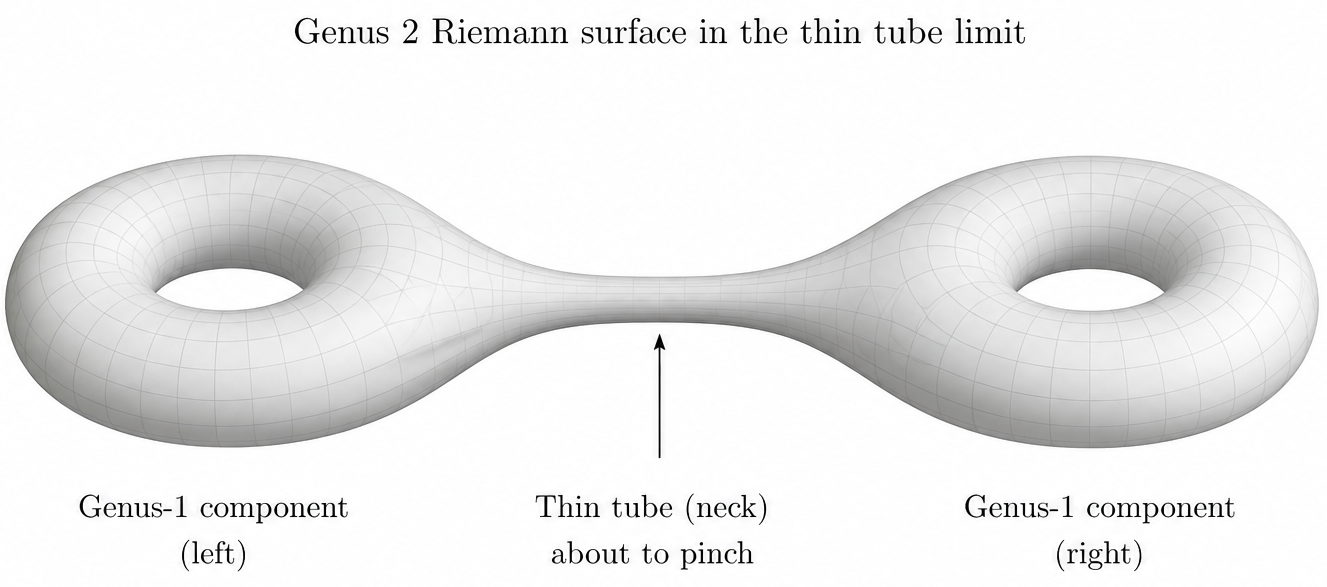}
    \caption{Genus-two surface in the separating degeneration breaks into two one-genus surfaces where the middle cycle is pinched off.}
    \label{fig:2}
\end{figure}

\noindent
To proceed with ECFTs, the main claim was that the existence of ECFTs with central charge $c=24k,\,\, k=1,2,3,\cdots\,,$ has a pure gravity dual in the bulk \cite{Yin:2007gv}. To begin with, the holomorphic part of the handlebody partition function at genus-two (The pictorial description of such genus-two handlebody is given in Fig.~\eqref{fig:handlebody}) for pure gravity is given by,
\begin{align}
    \mathbfcal{Z}_{h.b}=\sum_{\gamma\in \textrm{Map}(\Sigma_2)}\det(C\mathbf{\Omega}+D)^{-2k} Z_{\text{Saddle}}(k,\gamma\cdot \mathbf{\Omega})\label{2.1u}
\end{align}\\
where, the summand with loop-corrections are given by,
\begin{align}
    Z_{\text{Saddle}}(k, \mathbf{\Omega})=e^{kS_0(\mathbf\Omega)+S_1(\mathbf\Omega)+\frac{1}{k}S_2(\mathbf\Omega)+\cdots}\label{2.2u}
\end{align}
and $\gamma$ is the mapping class group (MCG) transformation for genus-two Riemann surfaces. \textcolor{black}{It is the group of all the gauge transformations modulo small gauge transformations}. As we will describe later, it is generated by the dehn-twists along the non-trivial cycles of the genus-$g$ Riemann surface. The MCG is represented by a larger set than of $Sp(2g,\mathbb{Z})$ transformations (for genus-$g$ surfaces) for $g\geq2$, while the \textit{period matrix} ($\mathbf{\Omega}$) for genus-$g$ Riemann surface is given by\,, \begin{equation}
    \mathbf\Omega=
    \begin{pmatrix}
     \Omega_{11} &\,\,\,\, \Omega_{12}\\
    \Omega_{12} &\,\,\,\,\Omega_{22}
    \end{pmatrix}
    \label{2.3r}
\end{equation}
alongwith $k^{1-\ell}S_{\ell}(\mathbf{\Omega})$ is the $\ell$-loop free-energy correction in the partition function in \eqref{2.1u}. Now, for $k=1$  ECFT, we have, by \cite{Yin:2007gv},
\begin{align}
    \begin{split}
    e^{S_0(\mathbf{\Omega})}=\frac{\mathbfcal{F}(\mathbf \Omega)^{12}}{\chi_{10}(\mathbf \Omega)};\,\,\,\,\,\,\,\,\,\,\,\,\,\, \chi_{10}(\mathbf \Omega)=\text{Igusa cusp form\,.}
    \end{split}
\end{align}
$\chi_{10}(\mathbf{\Omega})$ given by the following expression,
\begin{align}
  \chi_{10}(\mathbf{\Omega})=\prod_{(\delta,\epsilon)\,even}^{10}\theta^2[\delta,\epsilon](0|\mathbf{\Omega})\,, \quad
\end{align}
where, \begin{align}
    |\theta[\delta,\epsilon](z,\tau)|=\exp\bigg[-\pi \delta\cdot \text{Im}\tau\cdot \delta-2\pi \delta\cdot \text{Im} z\bigg]\underbrace{|\theta(z+\delta\cdot z+\epsilon|\tau)|}_{\exp[-i\pi \delta\cdot z\cdot\delta-2\pi i \delta\cdot z]\sum_{x\in \mathbb{Z}^g}\exp[\pi i x\cdot\tau\cdot x+2\pi i x\cdot z]}
\end{align}
Now, this is a Siegel modular form of weight (10,0) and the corresponding modular invariant norm is given by,
\begin{align}
    ||\chi_{10}(\mathbf{\Omega})||=2^{-12}(\textrm{det Im}(\mathbf{\Omega}))^5|\,\chi_{10}(\mathbf{\Omega})|\,.
\end{align}
Now, using the \textit{Schottky parametrization} $\mathbfcal{F}(\mathbf \Omega)$ is given by the following expression,
\begin{align}
    \mathbfcal{F}(\mathbf\Omega)=\prod_{\Gamma \,\textrm{prim}.}\prod_{m=1}^{\infty}(1-q^{m}_{\Gamma})
    \end{align}
whose solution in genus-two is given by \cite{Yin:2007gv} \footnote{where $\hat{E}_{\tau,n}$ denotes the $n$-th Eisenstein series $E_n(\tau)$ with the constant term removed,
and normalized so that,
\[
\hat{E}_{n,\rho} \;=\; \sum_{m=1}^\infty \frac{m^{\,n-1} q^m}{1 - q^m}\,, \quad q=e^{2\pi i\tau}\,.
\]
} ,
\begin{align}
\begin{split}
\frac{\mathbfcal{F}(\mathbf{\Omega})}{\prod_{m=1}^\infty (1 - q^m)^2 (1 - s^m)^2}
= &1
+ \frac{(2\pi i \Omega_{12})^2}{4} \, \hat{E}_{2,\Omega_{11}} \hat{E}_{2,\Omega_{22}}
+ \frac{(2\pi i \Omega_{12})^4}{3} \Big[
 -2 (\hat{E}_{2,\Omega_{11}})^2 \hat{E}_{2,\Omega_{22}}\\&
 -2\hat{E}_{2,\Omega_{11}} (\hat{E}_{2,\Omega_{22}})^{\,2}
 +48 (\hat{E}_{2,\Omega_{11}})^2 (\hat{E}_{2,\Omega_{22}})^2
 -10 (\hat{E}_{2,\Omega_{11}})^2 \hat{E}_{4,\Omega_{22}}\\&
 -10 (\hat{E}_{2,\Omega_{22}})^2 \hat{E}_{4,\Omega_{11}}
 -5 \hat{E}_{4,\Omega_{11}} \hat{E}_{4,\Omega_{22}}
\Big]+ O(\Omega_{12}^6)
\end{split}
\end{align}
where, $\hat{E}_k$ is the normalized Eisenstein series of modular weight $k$ and $q=e^{2\pi i\tau}\,.$
    For further details on Schottky parametrization in this context, we refer the reader to \cite{Yin:2007gv}. Therefore, combining everything, using \eqref{2.2u} and in the near \textit{separating degeneration} (as shown in Fig.~\ref{fig:2}) i.e $\Omega_{12}\rightarrow 0$ limit, the fake CFT partition function at genus two ($Z^{g=2}_{\text{fake}}$) is given by \cite{Yin:2007gv},
\begin{align}
    \begin{split}
 Z_{\text{Saddle}}(k, \mathbf{\Omega})=Z^{g=2}_{\text{fake}}=&G(\mathbf\Omega)^k\epsilon^{-2k}\Bigg[Z_{\textrm{vir}}(\tau_1)Z_{\textrm{vir}}(\tau_2)\\&-\frac{\epsilon^2}{48k\pi^2 }\partial_{\tau_1}Z_{\textrm{vir}}(\tau_1)\partial_{\tau_2}Z_{\textrm{vir}}(\tau_2)+\mathcal{O}(\epsilon^4)+\cdots\Bigg]\label{2.11u}
    \end{split}
\end{align}
\noindent
where, the modular parameters as the entries of the period matrix mentioned in \eqref{2.3r} is given by \cite{Tuite:1999id},
\begin{align}
\begin{split}
   & \Omega_{11}=\tau_1+\frac{\epsilon^2}{2\pi i }\hat E_2(q_2)+\cdots\\&
   \Omega_{22}=\tau_2+\frac{\epsilon^2}{2\pi i }\hat E_2(q_1)+\cdots\,,
   \\&
    \Omega_{12}=\frac{\epsilon}{2\pi i }(1+\hat{E}_2(q_1)\hat{E}_2(q_2)\epsilon^2+\cdots)\label{2.9k}
    \end{split}
\end{align}
with the following definition of the normalized Eisenstein series,
\begin{align}
\begin{split}
\hat{E}_{2l}(q) &:= -\frac{B_{2l}}{(2l)!} E_{2l}(q) = -\frac{B_{2l}}{(2l)!} + \frac{4l}{(2l)!} q + \mathcal{O}(q^2)\,, \\&\hspace{1cm}
E_{2l}(q) = 1 - \frac{4l}{B_{2l}} \sum_{n=1}^{\infty} \sigma_{2l-1}(n) q^n\,.
\label{eq:E2-def}\end{split}
\end{align}
$B_{2l}$ are the Bernoulli numbers along with $\sigma_{2l-1}(n)=\sum_{d|n}d^{2l-1}$ are the divisor functions \footnote{The generating function for the Bernoulli numbers is
\begin{align}
    \frac{t}{e^t-1}=\sum_{l=0}^\infty B_{2l}\frac{t^{2l}}{2l!}\,.
\end{align}}\,.
Also note that, while writing the (\ref{2.11u}), we have inverted the expressions in (\ref{2.9k}), especially the expression of $\Omega_{12}$ and rewritten everything in terms of the series expansion of the small parameter $\epsilon$ (pinching parameter) in the separating degeneration limit. Hence, we can get,
\begin{align}
    \begin{split}
        &\epsilon= \Omega_{12}\Big[1-( \Omega_{12})^2\frac{E_{2}(\Omega_{11})E_2(\Omega_{22})}{144}+\mathcal{O}( \Omega_{12}^4)+\cdots\Big]\,,\\&
       G(\mathbf\Omega)=1-\epsilon^2\frac{E_{2}(\tau_1)E_2(\tau_2)}{72}+\mathcal{O}(\epsilon^4)+\cdots\,.
    \end{split}
\end{align}
\noindent
At the leading order $\mathcal{O}\big(1/\epsilon^2\big)$, we have two copies of the decomposed genus-one partition function for separating degeneration as shown in Fig.~\eqref{fig:2}.
\noindent
The genus-one vacuum Virasoro character relevant for the genus-one partition function is given by,
\begin{align}
    Z_{\textrm{vir}}(\tau)=q^{-k}\prod_{n=2}^\infty(1-q^n)^{-1}=(1-q)\frac{q^{-k+1/24}}{\eta(q)}\,.\label{2.10k}
\end{align}
Now, the handlebody fake partition function in \eqref{2.11u} can be compactly written in the next-to-leading (in the leading order, the genus-two surface is the product of two genus-one surfaces) order in $\epsilon$ (using (\ref{2.10k})) as follows,
\begin{align}
    \begin{split}
 Z_{\text{fake}}^{g=2}(\{q_1,q_2\})=&\frac{1}{\epsilon^2}\, Z_{\textrm{vir}}(q_1)Z_{\textrm{vir}}(q_2)\\&\hspace{-2 cm}+\underbrace{\frac{1}{144\times 48}\Bigg(\frac{q_1^{-k+1/24}}{\eta(q_1)}\frac{q_2^{-k+1/24}}{\eta(q_2)}\Sigma(q_1)\Sigma(q_2)-\frac{\prod_{i=1}^2Z_{\textrm{vir}}(q_i)E_2(q_i)}{72}\Bigg)}_{\text{seed}}\times \text{Anti hol.}\\&\hspace{8 cm}+\text{Modular Images}\label{2.16y}
    \end{split}
\end{align}
with the following definitions,
\begin{align}
   \Sigma(q_i)= \Bigg[(1-q_i) \bigg(E_2(q_i)-(24k-1)\bigg)-q_i\Bigg]\,, \quad i=1,2\,.
\end{align}
Now, the seed term is at order $\mathcal{O}(\epsilon^0)$. The above seed expression is general in $k$ and we need to sum over the Mapping class group (MCG) to fix the modular images contribution at genus-two separating degeneration limit. In obtaining \eqref{2.16y} we have used the following differential,
\begin{align}
    -4\pi i\frac{d}{d\tau}\ln[\eta(\tau)]=G_2({\tau})\,, \,\,\,\,\,\,\,\,\,\,\,\,\,\,E_{2}(\tau)=\frac{G_{2}(\tau)}{2\zeta(2)}
\end{align}\\
where the Riemann zeta function is given by
\begin{align}
    \zeta(x):=\frac{1}{\Gamma(x)}\rmint_{0}^\infty \frac{u^{x-1} }{e^{u}-1}du\,.
\end{align}
\noindent
Now, we sum over modular images $\gamma$, rendering the partition function modular invariant at this specific order in the pinching parameter expansion. The $\mathcal{O}(1/\epsilon^2)$ part is easy, as it is just a product of two $g=1$ contributions and hence one needs to perform two independent sums over the $PSL(2,\mathbb{Z})$ parameters on each of the two $Z_{\textrm{vir}}$ factors. The sum over the modular image on the next term, i.e. the seed term appearing at $\mathcal{O}(\epsilon^0)$, is more involved, and we now discuss it in detail. First, we rewrite the seed term for generic $q_1$ and $q_2$ in the following way,\\
\begin{align}
\begin{split}
\text{seed}:=&\Bigg[\prod_{i=1}^2\frac{q_i^{-k+1/24}}{\eta(q_i)^2}\bigg((1-q_i) \bigg(E_2(q)-(24k-1)\bigg)-q_i\bigg)\\&\hspace{1 cm}-\frac{\prod_{i=1}^2Z_{\textrm{vir}}(q_i)E_2(q_i)}{72}\Bigg]\times\text{Anti hol.}+\underbrace{\text{Modular Images}}_{\text{summed over $Sp(4,\mathbb{Z})$ images}}\,.
\label{2.19i}
\end{split}
\end{align}
\noindent
Now, one needs to sum over the mapping class group of the genus-two handlebody to render the gravity partition function. Before performing the sum, we write down a general term of the sum. It can be casted as (for $\tau_1=\tau_2=\tau$),
\begin{align}
\begin{split}
     \mathbfcal{E}
  (n,m,s_1,w)=&\frac{1}{\Im(\tau)|\eta(\tau)|^4}\sum_{\gamma}{\Im(\tau)} \,(q)^{-n} (\bar q)^{-m}\\&\,\hspace{2 cm}\times\underbrace{E_2^{s_1}(\tau)E_{2}^w(\bar\tau)}_{\text{extra piece for genus-two handlebody}}\Bigg|_{\gamma};\forall s_1,w\in 0,1,2\label{3.17y}
  \end{split}
\end{align}
\noindent
\textcolor{black}{One should note that there is a qualitative difference of Eisenstein series terms in comparison to the genus-one computation of summing over the MCG.} Now, to perform the modular sum and seperate the inequivalent modular images, we compute the modular transformation of the summand itself (To cure overcounting of the same saddles). The modular tranformation is given by the following relations. The modular transformations for the Dedekind eta function is given by,
\begin{align}
    \begin{split}
        \eta\Bigg(\frac{a\tau+b}{c\tau+d}\Bigg)=\xi(a,b,c,d)(c\tau+d)^{1/2}\eta(\tau)
    \end{split}
\end{align}
 where the prefactor $\xi(a,b,c,d)$ is given by,

\begin{align}
    \xi(a,b,c,d):=e^{i\pi(\frac{a+d}{12c}-s(d,c)-\frac{1}{4})};\,\,\,\,\,\,\, s(d,c):=\sum_{n=1}^{c-1}\Bigg(\Bigg(\frac{n}{c}\Bigg)\Bigg)\Bigg(\Bigg(\frac{dn}{c}\Bigg)\Bigg)
\end{align}
$s(d,c)$ is known as the \textit{Dedekind sum}. Also the modular transformation of $G_2(\tau)$ is given by,
\begin{align}
    G_{2}\left(\frac{a\tau+b}{c\tau+d}\right)=(c\tau+d)^2G_{2}(\tau)+\cdots
\end{align}
Hence, the modular sum in \eqref{3.17y}  after the  generic modular transformation using the definitions of the elliptic nome, can be written as follows,


\begin{align}
    \mathbfcal{E} (n,m,s_1,w)=\frac{\,E_2^{s_1}(\tau)E_2^w(\bar \tau)}{|\eta(\tau)|^2}\sum_{c,d}|c\tau+d|^{{2s_1+2w-2}}\exp\left(2\pi \kappa \,\Im(\gamma\cdot\tau)+2\pi i \mu\, \Re(\gamma\cdot \tau)\right)\label{2.21e}
\end{align}
where $\kappa=m+n$ and $\mu=m-n$. A natural question is why, although the genus-two MCG is larger than $Sp(4,\mathbb{Z})$, the calculation uses $PSL(2,\mathbb{Z})\times PSL(2,\mathbb{Z})$. The reason is that we work near the separating degeneration, where the middle cycle becomes a thin tube and the genus-two Riemann surface splits into two tori. At the order considered here, the explicit calculation is therefore restricted to the block-diagonal subgroup $PSL(2,\mathbb{Z})\times PSL(2,\mathbb{Z})$ that preserves this degeneration; it is not the full $Sp(4,\mathbb{Z})$ modular sum. We return to this point in Section~\ref{sec3}.\\
Furthermore, for $\tau:=x'+iy'$ one can show that, \begin{align}
\begin{split}
&\Im(\mathbf{\gamma}\cdot \tau)=\frac{y'}{(cx'+d)^2+c^2y'^2}.\\&
\Re(\mathbf{\gamma}\cdot \tau)=\frac{a}{c}-\frac{cx'+d}{c((cx'+d)^2+c^2y^2)}.
\end{split}
\end{align}

\noindent
Considering, the whole fake partition function as the product of the seed operators, the exact form of the partition function at $k=2$, can be obtained using  \eqref{2.19i} explicitly for different elliptic nomes, $q_1$ and $q_2$.
\\Now, using the same description for the genus-two handlebody as of torus, and noting that the characters are perturbatively known in the expansion of the modular parameter  of the intermediate torus, one can systematically exploit modular invariance  at each order of the expansion. Before pursuing this analysis, however, it is
important to recall that the mapping class group (MCG) of genus-two Riemann
surfaces is given by extended $Sp(4,\mathbb{Z})$. In the gravitational context, the  MCG must be gauged after quantization. Accordingly, at the beginning of the next section we first review the structure of the genus-two MCG, and then
proceed to perform the sum over saddle points associated with the handlebody contribution, generated by the large diffeomorphisms.

\section{Mapping class group at genus-two}\label{sec3}
\begin{figure}[t!]
    \centering
\includegraphics[width=0.30\linewidth]{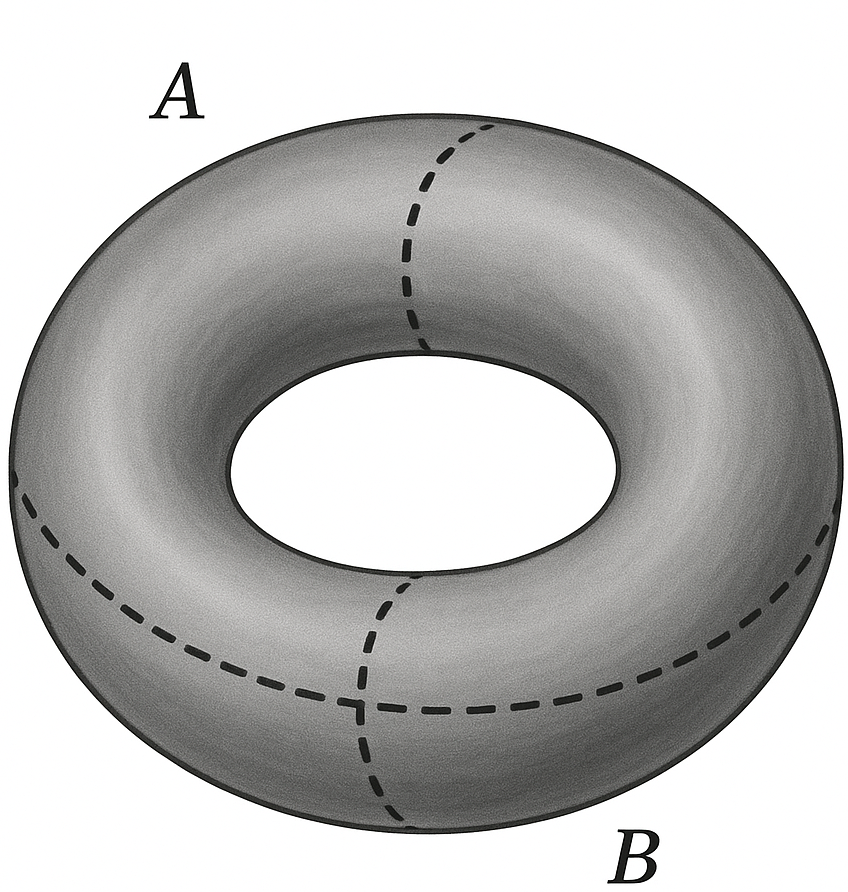}\,\,\,\,\,\hspace{2 cm}
\includegraphics[width=0.40\linewidth]{HAN.png}
    \caption{A genus one and a genus-two Riemann surface (handlebody). A genus-two Riemann surface can be prepared by sewing two one-punctured genus-one surfaces along the joining cycle. }
    \label{fig:41}
\end{figure}

Before turning to the explicit discussion of the genus-two modular sum, let us recall the structure of the mapping class group of a genus-two Riemann surface. As established in previous works, this group is isomorphic to the symplectic group $Sp(4,\mathbb{Z})$. In the present analysis, the detailed properties of genus-two conformal blocks are not required, since our primary concern is with the simplest three-dimensional bulk filling of the genus-two boundary surface. This role is played by the genus-two handlebody, which furnishes the natural and most elementary filling consistent with the action of the mapping class group. One should also note that here we work near the separating degeneration, where the handlebody contribution is the focus of our approximation. Pictorially, the genus-one $a$ cycle and $b$ cycles are shown in Fig.~\eqref{fig:41}. For the genus-two case the mapping class group and modular transformations are described below.
\subsection*{\textit{Genus-two mapping class group}}
The genus-two period matrix has a positive-definite imaginary part. The modular transformation matrix $\mathbfcal{M}$, which acts on the periods, preserves the canonical intersection form $\mathbfcal{M}^T J\mathbfcal{M}=J$, with \\
\begin{align}
    \mathbfcal{M}=\begin{pmatrix}
        A & B\\
        C & D
\end{pmatrix}\hspace{2 cm}J=\begin{pmatrix}
        0& -I_h \\
        I_h & 0
\end{pmatrix}\hspace{2 cm} \mathbfcal{M}\begin{pmatrix}
        A \\
        B
    \end{pmatrix}=\begin{pmatrix}
        \tilde{A} \\
        \tilde{B}
    \end{pmatrix}
\end{align}
\noindent
where, we assume that the entries $A,B,C,$ and $D$ are all $h\times h$ matrices. The modular action on the genus-two period matrix is more involved than in the genus-one case. The mapping class group is given by $\textbf{Map}(\Sigma_2)$, which is not same as $Sp(4,\mathbb{Z})$ and we will discuss the reason in this section at the end. The $Sp(4,\mathbb{Z})$ action on the period matrix is,
\begin{align}
\mathbf{\Omega}\rightarrow(A\mathbf{\Omega}+B)(C\mathbf{\Omega}+D)^{-1}.\label{4.30i}
\end{align}
where $\mathbf{\Omega}$ is the period matrix and $A,B,C,D$ are integer matrices satisfying $A^T D-C^T B=I_2$, $A^T C=C^T A$, and $B^T D=D^T B$.
Hence, after computing the partition function explicitly we discuss briefly the genus-two mapping class group. In addition to the two canonical pairs of cycles on the genus-two surface, we denote the separating cycle by $c_1$. At genus two, there are two \textit{holomorphic differentials}, and their periods are written as follows:\\
\begin{align}
\begin{split}
    &\rmint_{a_i} \omega_{j}=\delta_{ij}\,,\,\hspace{1 cm} \rmint_{b_i} \omega_{j}=\Omega_{ij}\,,\hspace{1 cm} \rmint_{c_1} \omega_{1}=1\,,\hspace{1 cm}\rmint_{c_1} \omega_{2}=-1.
    \end{split}
    \end{align}\\
    \noindent
\textbullet{\textbf{\,\,Modular}$\, \mathbf{Sp(4,\mathbb{Z})}$:} We now describe how the modular parameters transform under modular transformations. The genus-two period matrix is given by,
\begin{align}
    \mathbf{\Omega}=\begin{pmatrix}
    \Omega_{11}& \Omega_{12}\\
    \Omega_{12} &  \Omega_{22}
    \end{pmatrix}
\end{align}
where $\Omega_{ij}$'s in the leading order pinching limit of the middle torus of the genus-two surface, are defined in \eqref{2.9k}. In the pinching limit, the relevant modular action is generated by the left and right modular groups. \\

\noindent
We describe the actions of the generators of the left and right modular groups on the modular parameters as follows:\\

\noindent
\textbf{\textit{Action of the left modular group}}
The left modular group comprises $\langle S_1,T_1\rangle$ and similarly the right modular group.
Under the action of the left modular group the modular parameters change in the following way,
\begin{align}
    S_1:\tau_1\to-\frac{1}{\tau_1},\,\,\,\epsilon\to-\frac{\epsilon }{\tau_1}.
\end{align}
and $\Omega_{ij}$'s change according to,
\begin{align}
    \Omega_{11}\to -\frac{1}{\Omega_{11}},\,\,\,\,\,
    \Omega_{12}\to -\frac{\Omega_{12}}{\Omega_{11}},\,\,\,\,\,
    \Omega_{22}\to \Omega_{22}-\frac{\Omega_{12}^2}{\Omega_{11}}
\end{align}
 where, the action of $T_1$ is similar to \eqref{3.11r}. In the separating degeneration limit, we restrict attention not to the full $Sp(4,\mathbb{Z})$, but to its block-diagonal subgroup $PSL(2,\mathbb{Z})\times PSL(2,\mathbb{Z})$. Similar to what we have discussed above the right modular group action is given below.\\
 \noindent
\textbf{\textit{Action of the right modular group}}
 Similarly under the right modular group action generated by,$\langle S_2,T_2\rangle$:
\begin{align}
    S_2:\tau_2\to-\frac{1}{\tau_2},\,\,\,\epsilon\to-\frac{\epsilon }{\tau_2}.
\end{align}
and $\Omega_{ij}$'s change as \cite{Tuite:1999id},
\begin{align}
    \Omega_{22}\to -\frac{1}{\Omega_{22}},\,\,\,\,\,
    \Omega_{12}\to -\frac{\Omega_{12}}{\Omega_{22}},\,\,\,\,\,
    \Omega_{11}\to \Omega_{11}-\frac{\Omega_{12}^2}{\Omega_{22}}
\end{align}
Now after knowing the behaviour of the modular parameters under the modular transform, we proceed to effectively show the connection to $PSL(2,\mathbb{Z})$.\\

\noindent
\subsection{{Connection to diagonal subgroup}}
In this section, we describe the action of the block-diagonal $PSL(2,\mathbb{Z})\times PSL(2,\mathbb{Z})$ subgroup. Symbolically, we perform the following sum,
\begin{align}
Z_{\text{fake}}^{\text{mod}}:=\sum_{(\gamma_1,\gamma_2)\in PSL(2,\mathbb{Z})\times PSL(2,\mathbb{Z})}Z_{\text{fake}}^{g=2}(\gamma_1\cdot\tau_1,\gamma_2\cdot\tau_2)
\end{align}

\noindent
Here we describe how to perform the sum for modular-$S$ transform. The $PSL(2,\mathbb{Z})$ transformation matrix is given by, \begin{align}
    \gamma:=\begin{pmatrix}
    a & b\\
    c& d
\end{pmatrix}; \,\,\,\,\,\,ad-bc=1.
\end{align}
Translations $\tau \to \tau + 1$ are done by the matrix,
\begin{align}
   T =
\begin{pmatrix}
1 & 1 \\
0 & 1
\end{pmatrix}
\label{3.11r}
\end{align}
and its powers. Now we have the following,
\[
\gamma_{c,d,m,n} = T^n \cdot
\begin{pmatrix}
[d^{-1}]_c &[r]_{c,d} \\
c & d
\end{pmatrix}
\cdot T^m
= T^n \cdot \gamma_{c,d,0,0} \cdot T^m.
\]
For the above decomposition from now on we write the short notation of $\gamma_{c,d,0,0}$ replacing by $\gamma_{c,d}$.
Hence the set of $PSL(2,\mathbb{Z})$ elements without duplication are given by,
\begin{align}
PSL(2,\mathbb{Z}) = \left\{ T^n \right\}_{n\in \mathbb{Z}}
\cup
\left\{ T^n \cdot \gamma_{c,d} \cdot T^m
\;\middle|\;
c \geq 1,\; d \in (\mathbb{Z}/c\mathbb{Z})^*,\; m,n \in \mathbb{Z}
\right\}.
\end{align}
Hence, the modular transformation matrix $\gamma_{c,d}$ can be written as,
\begin{align}
    \gamma_{c,d}:=\begin{pmatrix}
        [d^{-1}]_c &[r]_{c,d} \\
c & d
    \end{pmatrix}
\end{align}
We also have $ad\equiv1 (\text{mod}  \,c)$ and a is a multiplicative inverse of $d$ and hence we can write $a$ as,
\begin{align}
    a=[d^{-1}]_{c}+cn ;\,\,\,\,\,\,\,\, \text{$n$ being an integer.}
\end{align}
One should note that to find the gravity partition function we need to sum over all the left cosets of $Sp(4,\mathbb{Z})$ modulo the stabilizer at the cusp ($\Gamma_{\infty}$). For a $Sp(4,\mathbb{Z})$ matrix $\gamma$ is generalised from its $PSL(2,\mathbb{Z})$ definition and is given by,
    \begin{align}
    \gamma:=\begin{pmatrix}
    A & B\\
    C& D
\end{pmatrix};
\end{align}
where, $A,B,C,D$ are matrices. satisfying the following properties,\begin{align}
    \gamma^{T}J\gamma=J, \qquad\,\,\,\,J=\begin{pmatrix}
    0 & I_2\\
    -I_2& 0
\end{pmatrix}.
\end{align}
with $I_2$ defining $2\times 2$ identity matrices. Now, the expression for the stabilizer at the cusp is given by,
\begin{align}
    \Gamma_{\infty}=\left\{\begin{pmatrix}
    A & B\\
    0& (A^{T})^{-1}
\end{pmatrix}\in Sp(4,\mathbb{Z})\right\}
\end{align}
It is also called the \textit{maximal parabolic subgroup} of $Sp(4,\mathbb{Z})$, that stabilizes the cusp at $i\infty$. Having all the ingredients in hand we perform the modular sum in appendix~\ref{appB}.
The result of the total partition function is given by,
\begin{align}
\begin{split}
     Z_{\textrm{grav}}^{g=2}(\tau_1=\tau_2=\tau)&:=  Z_{\textrm{fake}}^{g=2}(\tau_1=\tau_2=\tau)+\text{Modular images\,} (\mathbfcal{E}(\kappa,\mu,s_1,w))
     \end{split}
\end{align}
\noindent
Before commenting on the handlebody OPE density for pure gravity in $AdS_3$ at genus-two, we find the handlebody OPE density in ECFT using the direct inversion of the partition function. Instead of directly using the modular bootstrap at genus-two we perturbatively construct the genus-two density using the lower genus (genus-one) OPE data. Hence, before directly jumping to the genus-two data we primarily focus on genus-one OPE data and then build the OPE density of genus-two using that (lower genus) information in the later sections.
\subsection{Actual mapping class group is larger than $Sp(4,\mathbb{Z})$}
Before proceeding, in this subsection we comment on why the mapping class group at genus-two is larger than the $Sp(4,\mathbb{Z})$. 
Let the conformal boundary of the three-dimensional surface has a bulk which can be considered as closed Riemann surface of genus $g$ and is given by, $\Sigma_g.$ Choose a canonical basis of one-cycles$
\{a_1,\ldots,a_g,b_1,\ldots,b_g\},
$
with the following choice of the intersection pairing,
\begin{align}
    a_i\cap b_j=\delta_{ij},
\qquad
a_i\cap a_j=b_i\cap b_j=0.
\end{align}

\noindent
An \emph{empty handlebody} $H_g$ is the simplest three-manifold whose boundary is $\Sigma_g$. For a reference handlebody, one may choose the cycles $a_i$ to be contractible in the bulk. Equivalently $a_i$'s can be written as,
\begin{align}
    a_i=\partial D_i,
\end{align}
where each $D_i$ is a disk in the bulk handlebody. Thus a handlebody filling is not specified only by the abstract boundary surface. One must also specify which system of boundary cycles becomes contractible in the bulk. Different choices of contractible cycles can therefore correspond to different semiclassical gravitational saddles in the gravity theory. Now, the mapping class group of $\Sigma_g$ is given by,
\begin{align}
    \textbf{Map}(\Sigma_g)
=\frac{\operatorname{Diff}^+(\Sigma_g)}{\operatorname{Diff}_0(\Sigma_g)}
\end{align}
hence, it consists of orientation-preserving diffeomorphisms modulo those, which are continuously connected to the identity. Now, to describe why the $Sp(4,\mathbb{Z})$ is not enough, as the mapping class group of the genus-two surface, we introduce the concept of the \textit{Torelli subgroup}. It is given by the kernel of the following map,
\begin{align}
    \sigma:\textbf{Map}(\Sigma_2)\to Sp(4,\mathbb{Z}).
\end{align}
For genus-one surface the Torelli subgroup is trivial. But for higher genus ($g\geq2$) surfaces it is non-trivial due to the separating curves ($c_1$-cycle for genus-two). The genus-two surface has the first homology as\footnote{As it has four independent non-trivial cycles for Dehn twists generated by $a_1,a_2,b_1,b_2$ and under abelianization by the commutator the first fundamental group $\pi_1$ has four independent cycles.},
\begin{align}
    H_1(\Sigma_2,\mathbb{Z})\sim \mathbb{Z}^4.
\end{align}
\begin{figure}
    \centering
\includegraphics[width=0.48\linewidth]{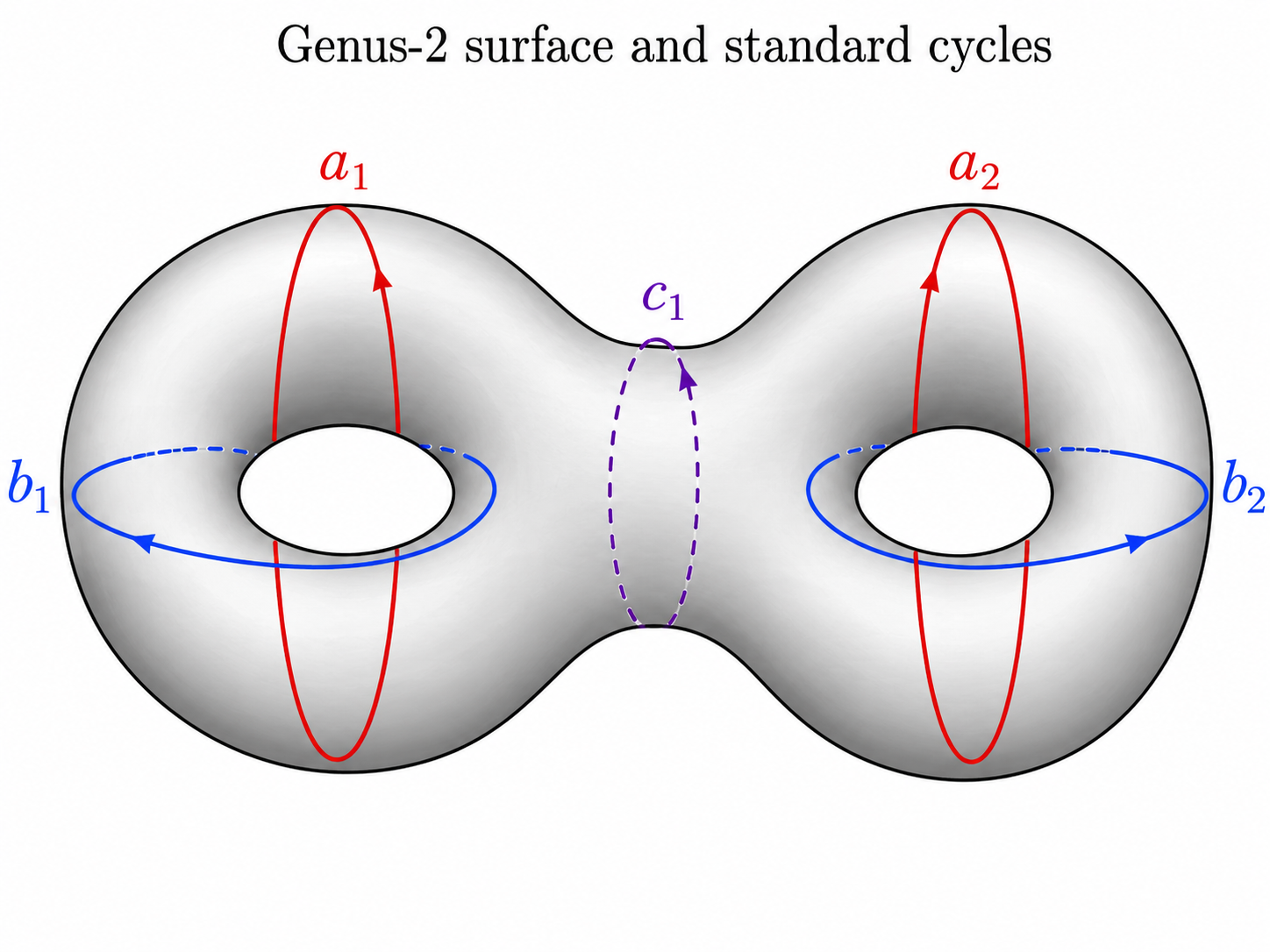}
    \caption{Picture depicting a genus-two surface with marked cycles. $c_1$ is the separating curve. If one pinches along $c_1$, as its a separating degeneration, the genus-two surface degenerates into two genus-one surfaces.}
    \label{fig:5}
\end{figure}
Now, the $c_1$-cycle has a separating curve rendering to $[c_1]=0$. Hence, it can be represented by $c_1=(0,0,0,0)^T$.
Therefore, it is possible to generate non-trivial large diffeomorphisms along the $c_1-$cycle as shown in Fig.~\eqref{fig:5}, which are not continuously connected to identity. Hence, it is a part of mapping class group. However, homologically it is trivial because,
\begin{align}
    \mathbfcal{T}_{c_{1}}[\gamma]=[c_1]+n[c_1]=[c_1]
\end{align}
where, $\mathbfcal{T}_{c_{1}}$ is the dehn-twist along the curve $c_1$. Though, $\mathbfcal{T}_{c_{1}}$ is a non trivial element, $\sigma\cdot[\mathbfcal{T}_{c_{1}}]$ is homologically trivial. Therefore, it is a non-trivial element of the mapping class group, but beyond the regime of $Sp(4,\mathbb{Z})$. Next, we proceed to comment on the genus-one density of state.

\vspace{0.2 cm}
\noindent
\textit{\textbf{Comments on genus-one density of states:}}
At genus one, the torus partition function can be decomposed into Virasoro characters and inverted to obtain the primary density in fixed-spin sectors. The MWK Poincar\'e construction gives a continuous density with the expected Cardy growth, but it can become negative in odd-spin sectors near the black-hole threshold \cite{Maloney:2007ud,Alday:2019vdr,Alday:2020qkm}. Since a compact genus-two crossing kernel is not presently available, we use the corresponding genus-one inversion kernel as the building block of the separating-channel construction. The detailed derivation is given in the next section.
\section{Fake partition function and OPE density in handlebody: relation to MWK}
\label{sec4}
The aim of this section is to identify the genus-two analogue of the MWK vacuum seed and then use it to extract the handlebody OPE density.  At genus one the MWK partition function is obtained by summing the vacuum contribution over modular images,
\begin{align}
    Z_{\textrm{MWK}}(\tau,\bar \tau)
=\sum_{\gamma\in\Gamma_\infty\backslash PSL(2,\mathbb Z)}
    Z_{0,1}(\gamma\cdot\tau,\gamma\cdot\bar\tau),
    \qquad
    Z_{0,1}(\gamma\cdot\tau,\gamma\cdot\bar\tau)
    =|\chi_0(\gamma\cdot\tau)|^2 .
\end{align}
The same modular image \(\gamma\) acts on both the holomorphic and anti-holomorphic factors.  Thus \(Z_{\textrm{MWK}}\) is not the product of two independent chiral Poincar\'e sums.  This distinction is important at genus two, because the gravity object we want is the same-saddle completion, not a holomorphically factorized ECFT partition function.

For an extremal CFT with $c=24k \,(\text{with} \, k\geq2)$, the holomorphic handlebody contribution is generated by the fake partition function,
\begin{align}
    \mathbfcal{Z}^{\textrm{hol}}_{\textrm{h.b.}}
    =
    \sum_{\gamma\in\Gamma_\infty\backslash Sp(4,\mathbb{Z})}
    \det(C\mathbf{\Omega}+D)^{-2k}
    Z_{\textrm{fake}}(k,\gamma\cdot \mathbf{\Omega}) .
\end{align}
Here \(Z_{\textrm{fake}}\) is the handlebody saddle seed.  The MWK-like non-chiral gravity completion is therefore given by,
\begin{align}
    Z^{\textrm{grav}}_{g=2}(\Omega,\bar\Omega)
    =
    \sum_{\gamma\in\Gamma_\infty\backslash Sp(4,\mathbb{Z})}
    |\det(C\Omega+D)|^{-4k}
    Z_{\textrm{fake}}(k,\gamma\cdot\Omega)\,
    \overline{Z_{\textrm{fake}}(k,\gamma\cdot\Omega)} .
    \label{eq:section5-same-saddle}
\end{align}
The genus-two MWK-type completion constitutes a more robust non-chiral (same saddle) completion than the original factorized chiral construction,
$\mathbfcal{Z}_{\textrm{h.b.}}^{\textrm{hol}}
\overline{\mathbfcal{Z}_{\textrm{h.b.}}^{\textrm{hol}}}.$ In the later case the sum has mixing between different saddles and has more information. Such a product would sum holomorphic and anti-holomorphic images independently, whereas the MWK construction keeps the same image in both sectors. Now, near the separating degeneration, the fake genus-two vacuum character reduces to two torus vacuum characters sewn by the pinching parameter, \(\epsilon\),
\begin{align}
    \chi_{0}^{(2)}(\boldsymbol{\Omega},k=1)
    =
    G(\mathbf\Omega)\epsilon^{-2}
    \left[
    \chi_{0}(\tau_1)\chi_{0}(\tau_2)
    -
    \frac{\epsilon^2}{48\pi^2}
    \partial_{\tau_1}\chi_{0}(\tau_1)
    \partial_{\tau_2}\chi_{0}(\tau_2)
    +\mathcal{O}(\epsilon^4)
    \right].
\end{align}
Non-vacuum characters have the same sewing structure. Their leading behaviour is given by,
\begin{align*}
    \epsilon^{h-2k}\chi_h(\tau_1)\chi_h(\tau_2).
\end{align*}
The corresponding subleading terms follow from the same sewing expansion.\footnote{For \(k=1\),
\[
\chi_{h}^{(2)}(\boldsymbol{\Omega},k=1)
=
\epsilon^{h-2k}G(\boldsymbol{\Omega})
\left[
\chi_h(\tau_1)\chi_h(\tau_2)
-\frac{\epsilon^2}{48\pi^2}
\partial_{\tau_1}\chi_h(\tau_1)\partial_{\tau_2}\chi_h(\tau_2)
+\mathcal O(\epsilon^4)
\right].
\]
For a general CFT, the genus-two partition function is a sum over genus-two conformal blocks weighted by \(|C_{ij\ell}|^2\).  This is why genus two probes OPE data rather than only the spectrum.} We first review the genus-one MWK density in a form that will be copied in the separating degeneration of genus-two calculation.  The review is included only to fix notation and to clarify where the primary density differs from the full density with descendants.
\subsection{Density of states for the MWK partition function: direct inversion}

In this section we derive the genus-one primary density associated with the MWK partition function by stripping off the Virasoro descendant contribution, equivalently the one-loop factor. We therefore define the reduced partition function as,
\begin{align}
    \mathcal Z_{\rm MWK}(\tau,\bar\tau)
    =\underbrace{ |\eta(\tau)|^2 }_{\text{one-loop factor}}Z_{\rm MWK}(\tau,\bar\tau).\label{4.5t}
\end{align}
Equation (4.5) expresses the MWK partition function as a sum over the modular images of the vacuum contribution. Here
$
\gamma=\begin{pmatrix}a&b\\ c&d\end{pmatrix}\in PSL(2,\mathbb Z),\, ad-bc=1,
$acts on the torus modulus as
\begin{align}
\gamma\tau=\frac{a\tau+b}{c\tau+d}.
\end{align}

\noindent
Now, Using the modular property,
\begin{align}
    \eta(\gamma\tau)
    = \nu(\gamma)(c\tau+d)^{1/2}\eta(\tau),
\end{align}
with, the transformation matrix given as,
\begin{align}
    \gamma=
    \begin{pmatrix}
        a & b \\
        c & d
    \end{pmatrix}
    \in PSL(2,\mathbb Z),
    \qquad
    \gamma\tau = \frac{a\tau+b}{c\tau+d},
\end{align}
we can express \eqref{4.5t} as,
\begin{align}
\begin{split}
    \mathcal Z_{\rm MWK}(\tau,\bar\tau)
    &=\sum_{\gamma\in \Gamma_\infty\backslash PSL(2,\mathbb Z)}
    (c\tau+d)^{-1/2}(c\bar\tau+d)^{-1/2}
    \,\mathcal Z_{\rm vac}(\gamma\tau,\gamma\bar\tau),
    \label{4.8h}
    \end{split}
    \end{align}
     where,
    \begin{align}
        \begin{split}
    \mathcal Z_{\rm vac}(\tau,\bar\tau)
    &=
    q^{-\hat c}\bar q^{-\hat c}(1-q)(1-\bar q)=
    \sum_{\ell,\bar\ell=0}^{1}
    (-1)^{\ell+\bar\ell}
    q^{-\hat c+\ell}
    \bar q^{-\hat c+\bar\ell}.
\end{split}
\label{4.9h}
\end{align}
The sum in \eqref{4.8h} is taken over the coset space, $\Gamma_\infty\backslash PSL(2,\mathbb Z),$
where, $(\Gamma_\infty)$ is the parabolic subgroup that leaves the cusp point in the fundamental domain of the genus-one surface $(\tau=i\infty)$ invariant, generated by the modular transformation $(T:\tau\mapsto\tau+1)$. Thus, elements that differ only by a $(T^n)$ transformation are identified, avoiding an overcounting of the same bulk saddles. Equivalently, the inequivalent modular images can be labelled by relatively prime integers $((c,d))$, with $(\gcd(c,d)=1)$ and $((c,d)\sim(-c,-d))$ in $(PSL(2,\mathbb Z))$. Each term therefore represents a distinct modular image of the vacuum virasoro character, or equivalently a distinct ($AdS_3$) solid-torus saddle.\\
Schematically \eqref{4.8h} defines the sum over modular images over the vacuum partition function at genus-one. The reduced primary density is defined through the Fourier-Laplace transform,
\begin{align}
    \mathcal Z_{\rm MWK}(\tau,\bar\tau)
    =\sum_{j\in\mathbb Z}
    \rmint de\,
    \rho^{\rm MWK}_{\rm prim}(e,j)
    \,e^{2\pi i jx-2\pi ey},
    \qquad
    \tau=x+iy.
\end{align}
Equivalently, the density can be extracted using the inverse transform
\begin{align}
    \rho^{\rm MWK}_{\rm prim}(e,j)
    =
    \frac{1}{2\pi i}
    \rmint_{\sigma-i\infty}^{\sigma+i\infty}dy\,
    e^{2\pi ey}
    \rmint_0^1dx\,
    e^{-2\pi i jx}
    \,\mathcal Z_{\rm MWK}(x+iy,x-iy).
    \label{4.11n}
\end{align}
Now, $\sigma$ is chosen in such a way such that all the poles of the integrand falls into the right of it. Now, Substituting the Poincar\'e series representation \eqref{4.9h} into the inversion formula \eqref{4.11n} and explicitly writing the elliptic nome in terms of the exponential, gives the following expression,
\begin{align}
\begin{split}
    \mathcal Z_{\rm MWK}(x+iy,x-iy)
    &=
    \sum_{\ell,\bar\ell=0}^{1}
    (-1)^{\ell+\bar\ell}
    \sum_{c=1}^{\infty}
    \sum_{\substack{d\in\mathbb Z\\(c,d)=1}}
    \frac{1}{\sqrt{D_{c,d}(x,y)}}
    \\
    &\qquad \times
    \exp\!\left[
        \frac{2\pi i aJ}{c}
        -
        \frac{2\pi i(cx+d)J}{cD_{c,d}(x,y)}
        -
        \frac{2\pi yE}{D_{c,d}(x,y)}
    \right],
\end{split}
\end{align}
where
\begin{align}
    D_{c,d}(x,y)
    =(cx+d)^2+c^2y^2,
\end{align}
and we introduced the shorthand
\begin{align}
    E \equiv 2\hat c-(\ell+\bar\ell),
    \qquad
    J \equiv \ell-\bar\ell.
\end{align}
Now, further on we rewrite the sum over $(c,d)$ in terms of the standard parametrization,
\begin{align}
    c=s,
    \qquad
    d=-r+sN,
\end{align}
with,
\begin{align}
    s\geq 1,
    \qquad
    0\leq r<s,
    \qquad
    (r,s)=1,
    \qquad
    N\in\mathbb Z.
\end{align}
In terms of these variables, the partition function becomes
\begin{align}
\begin{split}
    \mathcal Z_{\rm MWK}(x+iy,x-iy)
    &=
    \sum_{\ell,\bar\ell=0}^{1}
    (-1)^{\ell+\bar\ell}
    \sum_{s=1}^{\infty}
    \sum_{\substack{0\le r<s\\(r,s)=1}}
    \sum_{N\in\mathbb Z}
    \mathcal H_{r,s,N}(x,y),
\end{split}
\end{align}
where, the summand can be expressed as,
\begin{align}
\begin{split}
    \mathcal H_{r,s,N}(x,y)
    &=
    \frac{1}{\sqrt{D_{r,s}(x+N,y)}}
    \\
    &\qquad \times
    \exp\!\left[
        \frac{2\pi i aJ}{s}
        +
        \frac{2\pi i(r-s(x+N))J}{sD_{r,s}(x+N,y)}
        -
        \frac{2\pi yE}{D_{r,s}(x+N,y)}
    \right],\label{4.17u}
\end{split}
\end{align}
with
\begin{align}
    D_{r,s}(x,y)
    =(r-sx)^2+s^2y^2.
\end{align}
Substituting this expression into the inverse transform \eqref{4.11n}, we obtain the following expression of the reduced density,
\begin{align}
\begin{split}
    \rho^{\rm MWK}_{\rm prim}(e,j)
    &=
    \frac{1}{2\pi i}
    \sum_{\ell,\bar\ell=0}^{1}
    (-1)^{\ell+\bar\ell}
    \sum_{s=1}^{\infty}
    \sum_{\substack{0\le r<s\\(r,s)=1}}
    \rmint_{\sigma-i\infty}^{\sigma+i\infty}dy\,
    e^{2\pi ey}
    \sum_{N\in\mathbb Z}
    \rmint_0^1dx\,
    e^{-2\pi i jx}
    \mathcal H_{r,s,N}(x,y).
    \label{4.19i}
\end{split}
\end{align}
Now, from the definition of \eqref{4.17u}, it is very transparent that, the integrand depends on the combination, $x+N-\frac{r}{s}.$
Hence, it is therefore convenient to introduce the shifted variables as follows,
\begin{align}
    t=x+N-\frac{r}{s}.
\end{align}
After this change of variables in \eqref{4.19i}, the $x$-integral becomes
\begin{align}
\begin{split}
    &\sum_{N\in\mathbb Z}
    \rmint_0^1dx\,
    e^{-2\pi i jx}
    \mathcal H_{r,s,N}(x,y)
    \\  & =
    e^{\frac{2\pi i}{s}(aJ-jr)}
    \sum_{N\in\mathbb Z}
    \rmint_{N-r/s}^{N-r/s+1}dt
    \,
    \frac{e^{-2\pi ijt}}{s\sqrt{t^2+y^2}}
    \exp\!\left[
        -\frac{2\pi iJt}{s^2(t^2+y^2)}
        -\frac{2\pi yE}{s^2(t^2+y^2)}
    \right].
\end{split}
\end{align}
Using the identity,
\begin{align}
    \bigcup_{N\in\mathbb Z}[N+a,N+a+1]=\mathbb R,
    \qquad
    \forall a\in\mathbb R,
\end{align}
Thus, the original sum over integer translates is unfolded into an ordinary Fourier transform over the real line. After including the remaining inverse Laplace transform over $y$, the reduced density is expressed in terms of a two-dimensional integral over $t$ and $y$, with all dependence on the data $(r,s)$ appearing through an overall phase and the $s$-dependent kernel.
the sum over $N$ combines into a Fourier transform over the full real line. Now, incorporating the integral over $y$ we can write,
\begin{align}
\begin{split}
    &e^{\frac{2\pi i}{s}(aJ-jr)}
    \rmint_{\sigma-i\infty}^{\sigma+i\infty}dy\,
    e^{2\pi ey}
    \rmint_{-\infty}^{\infty}dt\,
    \frac{e^{-2\pi ijt}}{s\sqrt{t^2+y^2}}
    \exp\!\left[
        -\frac{2\pi iJt}{s^2(t^2+y^2)}
        -\frac{2\pi yE}{s^2(t^2+y^2)}
    \right].
\end{split}
\end{align}
For performing the remaining $y$ integral, now introduce the variables
\begin{align}
    y-it = \frac{\beta_L}{2\pi},
    \qquad
    y+it = \frac{\beta_R}{2\pi},
\end{align}
in terms of which the exponent is diagonal. Now with the following introduction of new variables,
\begin{align}
    u=\frac{e+j}{2},
    \qquad
    v=\frac{e-j}{2},
    \qquad
    A=E+J,
    \qquad
    B=E-J,
\end{align}
we can finally express the result of the integral as follows,\\
\begin{align}
\begin{split}
    &\frac{1}{2s}
    \left(\frac{1}{2\pi i}\right)^2
    \rmint_{2\pi\sigma-i\infty}^{2\pi\sigma+i\infty}
    d\beta_L\,d\beta_R
    \,
    \frac{1}{\sqrt{\beta_L\beta_R}}
    \exp\!\left[
        u\beta_L
        +
        v\beta_R
        -
        \frac{4\pi^2}{s^2}
        \left(
            \frac{A}{\beta_L}
            +
            \frac{B}{\beta_R}
        \right)
    \right]
    \\
    &\qquad =
    \frac{1}{2\pi s}
    \frac{\cosh\!\left(4\pi\sqrt{-\frac{Au}{s^2}}\right)}{\sqrt{u}}
    \frac{\cosh\!\left(4\pi\sqrt{-\frac{Bv}{s^2}}\right)}{\sqrt{v}}
    \Theta(u)\Theta(v).
\end{split}
\end{align}\\
where, $\Theta(u/v)$ defines the standard Heaviside step function. Finally, the reduced primary density defined in \eqref{4.19i} takes the form,
\begin{align}
\begin{split}
    \rho^{\rm MWK}_{\rm prim}(e,j)
    &=
    \frac{1}{4\pi\sqrt{e^2-j^2}}
    \sum_{\ell,\bar\ell=0}^{1}
    (-1)^{\ell+\bar\ell}
    \sum_{s=1}^{\infty}
    \frac{K(j,\ell-\bar\ell;s)}{s}
    \cosh\!\left[
        \frac{4\pi}{\sqrt{2}\,s}
        \sqrt{e+j}
        \,\sqrt{\hat c-\ell}
    \right]
    \\
    &\qquad \times
    \cosh\!\left[
        \frac{4\pi}{\sqrt{2}\,s}
        \sqrt{e-j}
        \,\sqrt{\hat c-\bar\ell}
    \right],\label{4.29kkk}
\end{split}
\end{align}
where the Kloosterman sum is given by,
\begin{align}
    K(j,m;s)
    = \sum_{\substack{0\le r<s\\(r,s)=1}}
    \exp\!\left[
        \frac{2\pi i}{s}(am-jr)
    \right]
\end{align}
is the corresponding \textit{Kloosterman sum}.  This is the density for Virasoro primaries. The MWK density given in \eqref{4.29kkk} matches exactly with the modular bootstrap result in \cite{Benjamin:2019stq}.  If one instead inverts the unstripped MWK partition function, the Virasoro descendants must be included as the contribution, which leads to the total density including the descendants as follows,
\begin{align}
    \rho^{\rm MWK}_{\rm full}(e,j)
    =
    \sum_{N,\bar N\ge0}p(N)p(\bar N)\,
    \rho^{\rm MWK}_{\rm prim}
    \left(e-N-\bar N+\frac{1}{12},j-N+\bar N\right),
    \label{eq:mwk-full-desc-convolution}
\end{align}
where, $p(N)$ is the partition number. We will now apply the same logic to the genus-two handlebody in separating degeneration limit. Now, in the next section we proceed to compute the genus-two OPE density, using the data found for the genus-one scenario.

\section{Constructing the genus-two OPE density}\label{secc5}

We now construct the density in the separating channel. The logic has three steps.  First, the genus-two handlebody is written as a modular completion of the fake seed.  Second, near the separating degeneration this completion reduces to two torus modular sums, corrected order by order in the pinching parameter.  Third, after stripping the universal Virasoro descendants, the Fourier-Laplace coefficients give the reduced OPE density.
The formal holomorphic handlebody completion is given above in \eqref{2.1u}.
For the computation of gravity density, we propose the corresponding same-saddle, MWK-type completion given by,
\begin{align}
    Z_{\rm grav}^{g=2}(\Omega,\bar\Omega)
:=\sum_{\gamma\in\Gamma_\infty\backslash Sp(4,\mathbb Z)}
    \left|\det(C\Omega+D)\right|^{-4k}
    Z_{\rm fake}(k,\gamma\cdot\Omega)\,
    \overline{Z_{\rm fake}(k,\gamma\cdot\Omega)}.
    \label{eq:same-saddle-gravity}
\end{align}

\noindent
This is the formal genus-two analogue of MWK: the same modular image appears in the two chiral sectors. An independently factorized holomorphic $\times$ anti-holomorphic sum would define a different object.\\
The Poincar\'e sum should be viewed as a sum over distinct modular images of one seed saddle.  The quotient by $\Gamma_\infty$ removes transformations that leave the seed partition function unchanged, while the same-saddle prescription correlates the holomorphic and anti-holomorphic parts of each saddle.  This correlation is essential: summing the two chiral sectors independently would include pairs that do not arise from a single bulk saddle, which intrinsically has more information .

\noindent
For the explicit calculation, we now specialize to the separating degeneration and evaluate the block-diagonal subgroup contribution. We write
\begin{align}
    \Omega_{11}=\tau_1+\mathcal O(\epsilon^2),\qquad
    \Omega_{22}=\tau_2+\mathcal O(\epsilon^2),\qquad
    \Omega_{12}=\nu=\frac{\epsilon}{2\pi i}+\mathcal O(\epsilon^3).
\end{align}
In this limit the relevant modular action is the block-diagonal subgroup
\begin{align*}
    PSL(2,\mathbb Z)\times PSL(2,\mathbb Z),
\end{align*}
\,\,with the following, 
\begin{align}
    \gamma_i=
    \begin{pmatrix}
        a_i & b_i\\
        c_i & d_i
    \end{pmatrix},
    \qquad
    \alpha_i:=c_i\tau_i+d_i,
    \qquad
    \Delta_\gamma:=\alpha_1\alpha_2.
\end{align}
The sewing parameter transforms as,
\begin{align}
    \det(C\Omega+D)=\Delta_\gamma+\mathcal O(\epsilon^2),
    \qquad
    \epsilon_\gamma=\frac{\epsilon}{\Delta_\gamma}+\mathcal O(\epsilon^3).
    \label{eq:epsilon-transform}
\end{align}
The fake seed has the expansion,
\begin{align}
    Z_{\rm fake}(k,\Omega)
    =
    \epsilon^{-2k}
    \sum_{n\ge0}\epsilon^{2n}F_n(\tau_1,\tau_2).
    \label{eq:fake-separating-expansion}
\end{align}
For $k=1$, Eq.~\eqref{2.11u} gives,
\begin{align}
\begin{split}
    F_0(\tau_1,\tau_2)
    &=Z_{\rm vir}(\tau_1)Z_{\rm vir}(\tau_2),\\
    F_1(\tau_1,\tau_2)
    &=-\frac{1}{48\pi^2}
    \partial_{\tau_1}Z_{\rm vir}(\tau_1)
    \partial_{\tau_2}Z_{\rm vir}(\tau_2)
    -\frac{1}{72}E_2(\tau_1)E_2(\tau_2)
    Z_{\rm vir}(\tau_1)Z_{\rm vir}(\tau_2).
\end{split}
\end{align}
For the density calculation below, we specialize to $k=2$, so $c=48$, $\hat c=(c-1)/24=47/24$, and $A_0=-47/24$, and use the explicit $k=2$ seed polynomial given in Eq.~\eqref{eq:g2-F1-finite}; no general-$k$ extrapolation of $F_1$ is assumed.\\

\noindent
Substituting the separating expansion into the modular image gives,
\begin{align}
    \det(C\Omega+D)^{-2k}
    \epsilon_\gamma^{-2k+2n}
    F_n(\gamma_1\cdot\tau_1,\gamma_2\cdot\tau_2)
    =
    \epsilon^{-2k+2n}
    \Delta_\gamma^{-2n}
    F_n(\gamma_1\cdot\tau_1,\gamma_2\cdot\tau_2).
    \label{eq:det-sewing-cancellation}
\end{align}
Thus the determinant cancels the leading sewing pole, and the \(n\)-th order term keeps only the residual weight \(\Delta_\gamma^{-2n}\).  The same-saddle expansion can therefore be written as
\begin{align}
    Z_{\rm grav}^{g=2}(\Omega,\bar\Omega)
    =
    |\epsilon|^{-4k}
    \sum_{n,m\ge0}
    \epsilon^{2n}\bar\epsilon^{2m}
    Z_{n,m}(\tau_i,\bar\tau_i),
    \label{eq:Zgrav-separating}
\end{align}
with
\begin{align}
    Z_{n,m}(\tau_i,\bar\tau_i)
    =
    \sum_{\gamma\in\Gamma_{\infty}\backslash[PSL(2,\mathbb Z)\times PSL(2,\mathbb Z)]}
    \Delta_\gamma^{-2n}\bar\Delta_\gamma^{-2m}
    F_n(\gamma_1\cdot\tau_1,\gamma_2\cdot\tau_2)\,
    \bar F_m(\gamma_1\cdot\bar\tau_1,\gamma_2\cdot\bar\tau_2).
    \label{eq:Znm-def}
\end{align}
For the formal full completion, the object \(Z_{\rm grav}^{g=2}\) is modular covariant rather than scalar invariant:
\begin{align}
    Z_{\rm grav}^{g=2}(\gamma\cdot\Omega,\gamma\cdot\bar\Omega)
    =
    \left|\det(C\Omega+D)\right|^{4k}
    Z_{\rm grav}^{g=2}(\Omega,\bar\Omega).
    \label{eq:Zgrav-covariant}
\end{align}
A modular invariant is obtained by multiplying by \((\det\operatorname{Im}\Omega)^{2k}\):
\begin{align}
    \mathcal Z_{\rm grav}^{g=2}(\Omega,\bar\Omega)
    :=
    (\det\operatorname{Im}\Omega)^{2k}
    Z_{\rm grav}^{g=2}(\Omega,\bar\Omega).
    \label{eq:scalar-invariant}
\end{align}
The distinction between covariance and invariance is only a matter of modular weight.  The factor $(\det\operatorname{Im}\Omega)^{2k}$ supplies the opposite weight and therefore converts the holomorphic normalization used in the calculation into a scalar modular invariant object.  Near the separating cusp this factor is a finite polynomial in $\epsilon$ and $\bar\epsilon$, so it reshuffles nearby sewing orders but does not alter the inversion kernel derived below.

The determinant factor only modifies the density by a finite order-mixing operation in the pinching expansion.\footnote{Near the separating cusp,
\[
\det\operatorname{Im}\Omega
=
y_1y_2-\frac{(\epsilon+\bar\epsilon)^2}{16\pi^2}+\cdots,
\]
and therefore
\[
(\det\operatorname{Im}\Omega)^{2k}
=
\sum_{r=0}^{2k}
(-1)^r\binom{2k}{r}
(y_1y_2)^{2k-r}
\frac{(\epsilon+\bar\epsilon)^{2r}}{(4\pi)^{2r}}
+\cdots.
\]
This is the content of the scalar-invariant correction \(\eqref{eq:detY-expansion}\).  Since it only mixes finitely many neighboring orders, we first compute the reduced density associated with \(Z_{n,m}\).}
\begin{align}
    (\det\operatorname{Im}\Omega)^{2k}
    =
    \sum_{r=0}^{2k}
    (-1)^r
    \binom{2k}{r}
    (y_1y_2)^{2k-r}
    \frac{(\epsilon+\bar\epsilon)^{2r}}{(4\pi)^{2r}}
    +\cdots .
    \label{eq:detY-expansion}
\end{align}
We now strip the two genus-one descendant factors and define the following,
\begin{align}
    \mathcal A_{n,m}(\tau_i,\bar\tau_i)
    :=
    \prod_{i=1}^{2}|\eta(\tau_i)|^2
    Z_{n,m}(\tau_i,\bar\tau_i).
    \label{eq:A-def}
\end{align}
The reduced density is the Fourier-Laplace coefficient of this object:
\begin{align}
    \mathcal A_{n,m}
    =
    \sum_{j_1,j_2\in\mathbb Z}
    \rmint de_1\,de_2\,
    \rho_{n,m}(e_1,j_1;e_2,j_2)
    \prod_{i=1}^{2}
    e^{2\pi i j_i x_i}e^{-2\pi e_i y_i},
    \qquad
    \tau_i=x_i+iy_i.
    \label{eq:A-density-expansion}
\end{align}
Equivalently,
\begin{align}
    \rho_{n,m}(e_1,j_1;e_2,j_2)
    =\prod_{i=1}^{2}
    \left[
    \frac{1}{2\pi i}
    \rmint_{\sigma_i-i\infty}^{\sigma_i+i\infty}
    dy_i\,e^{2\pi e_i y_i}
    \rmint_0^1dx_i\,e^{-2\pi i j_i x_i}
    \right]
    \mathcal A_{n,m}.
    \label{eq:rho-inversion}
\end{align}
Equation~\eqref{eq:rho-inversion} can be understood directly from the spectral decomposition. For each torus, the integration over $x_i$ transforms (the partition function) the states with a definite integer spin $j_i$, whereas the inverse Laplace transform in $y_i$ picks out states with fixed energy $e_i$. The factor $|\eta(\tau_i)|^2$ strips off the universal contribution coming from Virasoro descendants. Consequently, $\rho_{n,m}$ captures the OPE data associated with primary states only, rather than the complete spectrum including descendants. One could, in principle, keep the descendant contribution explicitly. Doing so, however, makes the inversion substantially more involved, since it introduces additional sums over partition numbers, as described in the previous section.

\noindent
The stripped seed coefficients are defined by
\begin{align}
\begin{split}
    \eta(\tau_1)\eta(\tau_2)F_n(\tau_1,\tau_2)
    &=
    \sum_{\ell_1,\ell_2\ge0}
    f_n(\ell_1,\ell_2)\,
    q_1^{-\hat c+\ell_1}
    q_2^{-\hat c+\ell_2},\\
    \bar\eta(\bar\tau_1)\bar\eta(\bar\tau_2)\bar F_m(\bar\tau_1,\bar\tau_2)
    &=
    \sum_{\bar\ell_1,\bar\ell_2\ge0}
    \bar f_m(\bar\ell_1,\bar\ell_2)\,
    \bar q_1^{-\hat c+\bar\ell_1}
    \bar q_2^{-\hat c+\bar\ell_2}.
    \label{eq:fmbar-def}
\end{split}
\end{align}
For each torus, set
\begin{align}
    A_i=-\hat c+\ell_i,\qquad
    B_i=-\hat c+\bar\ell_i,\qquad
    E_i=A_i+B_i,\qquad
    J_i=A_i-B_i.
    \label{eq:EJ-def}
\end{align}
Then the two tori factorize after the inverse transform:
\begin{align}
    \begin{aligned}
    \rho_{n,m}(e_1,j_1;e_2,j_2)
    &=
    \sum_{\ell_1,\ell_2}
    \sum_{\bar\ell_1,\bar\ell_2}
    f_n(\ell_1,\ell_2)\bar f_m(\bar\ell_1,\bar\ell_2)
    \prod_{i=1}^{2}
    R_{n,m}
    \left(
    e_i,j_i\mid
    -\hat c+\ell_i,\,
    -\hat c+\bar\ell_i
    \right).
    \end{aligned}
    \label{eq:rho-master}
\end{align}
Here \(R_{n,m}\) is the one-torus Rademacher kernel contributing to the \((n,m)\) order correction. The kernel is given by the following expression,
\begin{align}
    R_{n,m}(e,j\mid A,B)
    =
    \sum_{s=1}^{\infty}
    K(j,A-B;s)\,
    s^{-1-2n-2m}\,
    R_{n,m}^{(1)}
    \left(
    e,j\mid
    \frac{A}{s^2},\frac{B}{s^2}
    \right).
    \label{eq:Rnm}
\end{align}
where, $K(j,J;s)$ be the Kloosterman sum,
\begin{align}
    K(j,J;s)
    :=\sum_{\substack{0\le r<s\\(r,s)=1}}\exp\left[
    2\pi i\left(-\frac{rj}{s}+\frac{a(r,s)J}{s}
    \right)
    \right],
    \qquad
    a(r,s)r\equiv -1\pmod s .\label{6.22r}
\end{align}
and the primitive local kernel is given by,
\begin{align}
    \begin{aligned}
    R_{n,m}^{(1)}(e,j\mid A,B)
    &=
    \frac{1}{2\pi i}
    \rmint_{\sigma-i\infty}^{\sigma+i\infty}
    dy\,e^{2\pi e y}
    \rmint_{-\infty}^{\infty} dx\,
    e^{-2\pi i jx}
    (x+iy)^{-\frac12-2n}
    (x-iy)^{-\frac12-2m}
    \\
    &\quad\times
    \exp\left[
    -\frac{2\pi ix(A-B)}{x^2+y^2}
    -
    \frac{2\pi y(A+B)}{x^2+y^2}
    \right].
    \end{aligned}
    \label{eq:Rlocal}
\end{align}
The master formula divides the problem into two pieces.  The coefficients $f_n\bar f_m$ contain the polar seed data, whereas $R_{n,m}$ transforms each polar term to the heavy spectrum required by modular covariance.  The integer $s$ labels inequivalent modular images (or saddles): the Kloosterman sum combines their phases, and the factor $s^{-1-2n-2m}$ increasingly suppresses large-$s$ images at higher sewing order.
For \(j<0\) one uses the reflection
\((j,J,n,m)\mapsto(-j,-J,m,n)\), while the \(j=0\) sector requires the usual regularized prescription. Equivalently,
\begin{align}
    R_{n,m}^{(1)}(e,j\mid A,B)
    =
    \frac{(-1)^{n+m}}{(4\pi^2)^{n+m}}
    \partial_A^{2n}\partial_B^{2m}
    R_{0,0}^{(1)}(e,j\mid A,B).
    \label{eq:R-derivative-form}
\end{align}
 Moreover, the derivative representation is useful because no new integral is needed at every order.  Differentiating with respect to the left and right polar energies $A$ and $B$ generates the extra denominator powers associated with $\epsilon^{2n}\bar\epsilon^{2m}$.  All higher kernels can therefore be derived from the leading kernel, provided that the derivatives are taken before the rescaling $A,B\mapsto A/s^2,B/s^2$ and the same regularization scheme is kept.\\ \par

\noindent
\textbf{Leading contribution.} For $(n=m=0)$ and $j\neq0$, the kernel for a fixed polar pair $(A,B)$ is given by,
\begin{align}
\begin{split}
    R_{0,0}(e,j\mid A,B)
    &=
    \frac{1}{4\pi\sqrt{e^2-j^2}}
    \sum_{s=1}^{\infty}
    \frac{K(j,A-B;s)}{s}
    \cosh\!\left[
        \frac{4\pi}{\sqrt{2}\,s}
        \sqrt{e+j}\sqrt{-A}
    \right]\\
    &\qquad\times
    \cosh\!\left[
        \frac{4\pi}{\sqrt{2}\,s}
        \sqrt{e-j}\sqrt{-B}
    \right],
\end{split}
\label{6.26t}
\end{align}
with the appropriate analytic continuation when $A$ or $B$ is non-negative. The $j=0$ sector requires a separate regularized prescription.

\noindent
Since $F_0=Z_{\rm vir}(\tau_1)Z_{\rm vir}(\tau_2)$ and
$\eta(\tau)Z_{\rm vir}(\tau)=(1-q)q^{-\hat c}$, the torus vacuum combination is,
\begin{align}
    \rho_0(e,j)
    =
    \sum_{\ell,\bar\ell=0}^{1}
    (-1)^{\ell+\bar\ell}
    R_{0,0}\!\left(
    e,j\mid-\hat c+\ell,-\hat c+\bar\ell
    \right).
    \label{eq:rho0-fourterms}
\end{align}
Hence, the leading genus-two density therefore factorizes as,
\begin{align}
    \rho_{0,0}(e_1,j_1;e_2,j_2)
    =\rho_0(e_1,j_1)\rho_0(e_2,j_2).
    \label{eq:rho00-factorized}
\end{align}
This factorization implies that any sign problem present in a genus-one factor is inherited by the leading separating contribution. At this pinching limit the thin tube (i.e the $\epsilon$-cycle) carries only the vacuum sewing contribution, so the two handles do not yet exchange nontrivial data.  This explains the factorization of density \eqref{eq:rho00-factorized} at leading order.  It also gives an immediate diagnostic of unitarity: if one torus factor is negative while the other is positive, their genus-two product is negative, and increasing the genus alone cannot remove the genus-one pathology at this order. Therefore it can be inferred that perturbatively the pathology may not be cured. We will be more precise about those issues next. \\ \par
\textbf{First correction.}
For $k=2$, the first correction is controlled by $F_1$. After stripping the eta factors,
\begin{align}
\eta(\tau_1)\eta(\tau_2)F_1(\tau_1,\tau_2)
	    =  \frac{1}{144\cdot96}
	    \sum_{\alpha\in\mathfrak P_2}
	    C_\alpha\,
	    q_1^{A_0+r_{\alpha 1}}
	    q_2^{A_0+r_{\alpha 2}}
	    E_2(q_1)^{s_{\alpha 1}}
	    E_2(q_2)^{s_{\alpha 2}}.
	    \label{eq:g2-F1-finite}
\end{align}
Although \eqref{eq:g2-F1-finite} looks compact, it is just a formal expression.  Each label $\alpha$ specifies a monomial, the shifts $r_{\alpha i}$ move the energies on the two tori, and the exponents $s_{\alpha i}$ determine whether an $E_2$ series is present or not.  Expanding that series introduces the non-negative integers $u_i$, after which every term can be inverted with the same one-torus kernel.
For compactness, define $\mathcal N_1=(144\cdot96)^{-1}$.
		Here \(A_0=-\hat c=-47/24\), and the finite set \(\mathfrak P_2\) contains the data for the \(k=2\) polynomial \(F_1\).\footnote{The tuples \((C_\alpha,r_{\alpha1},r_{\alpha2},s_{\alpha1},s_{\alpha2})\) are
		\begin{align}
		\begin{split}
		\mathfrak P_2=\{&
	(385,0,0,1,1),(-385,1,0,1,1),(-385,0,1,1,1),(385,1,1,1,1),
	(47,0,0,1,0),(-47,1,0,1,0)\\&,(-23,0,1,1,0),(23,1,1,1,0),
	(47,0,0,0,1),(-23,1,0,0,1),(-47,0,1,0,1),(23,1,1,0,1),\\ &
		(2209,0,0,0,0),(-1081,1,0,0,0),(-1081,0,1,0,0),(529,1,1,0,0)\}.
		\end{split}
		\end{align}
		} With
\begin{align}
	    E_2(q)^s=\sum_{u\ge 0} c_s(u)q^u,
	    \qquad
    c_0(u)=\delta_{u,0},
    \qquad
    c_1(0)=1,\qquad
    c_1(u)=-24\sigma_1(u)\quad(u>0),
	\end{align}
	the inverse transform gives,
\begin{align}
    \mathcal{G}_{k=2}(e_1,j_1,e_2,j_2;\epsilon,\bar\epsilon)
    =|\epsilon|^{-8}\left[
    \rho_{0,0}
    +\epsilon^2\rho_{1,0}
    +\bar\epsilon^2\rho_{0,1}
    +\mathcal{O}(\epsilon^4,\epsilon^2\bar\epsilon^2,\bar\epsilon^4)
    \right].
\end{align}
\noindent
The two linear corrections are given by,
\begin{align}
\begin{split}
    \rho_{1,0}(e_1,j_1;e_2,j_2)
    &=
    \mathcal N_1
    \sum_{\alpha\in\mathfrak P_2}
    \sum_{\bar\ell_1,\bar\ell_2=0}^{1}
    C_\alpha(-1)^{\bar\ell_1+\bar\ell_2}
    \sum_{u_1,u_2\ge0}
    \prod_{i=1}^2 c_{s_{\alpha i}}(u_i)\\&\hspace{3 cm}\times
    \prod_{i=1}^2
    R_{1,0}
    \left(
    e_i,j_i\Big|
    A_0+r_{\alpha i}+u_i,\,
    A_0+\bar\ell_i
    \right),
\end{split}
\end{align}
\begin{align}
\begin{split}
    \rho_{0,1}(e_1,j_1;e_2,j_2)
    &=
    \mathcal N_1
    \sum_{\beta\in\mathfrak P_2}
    \sum_{\ell_1,\ell_2=0}^{1}
    \bar C_\beta(-1)^{\ell_1+\ell_2}
    \sum_{v_1,v_2\ge0}
    \prod_{i=1}^2 c_{s_{\beta i}}(v_i)\\
    &\quad\hspace{3 cm}\times
    \prod_{i=1}^2
    R_{0,1}
    \left(
    e_i,j_i\Big|
    A_0+\ell_i,\,
    A_0+r_{\beta i}+v_i
    \right).
\end{split}
\label{5.30jj}
\end{align}
The kernels are derivatives of the primitive leading kernel,
\begin{align}
\begin{split}
    R_{1,0}(e,j|A,B)
    &=
    -\frac{1}{4\pi^2}
    \sum_{s=1}^{\infty}
    K(j,A-B;s)\,s^{-3}
    \left.
    \partial_{\mathcal A}^2
    R_{0,0}^{(1)}(e,j|\mathcal A,\mathcal B)
    \right|_{\mathcal A=A/s^2,\,\mathcal B=B/s^2},
\\
    R_{0,1}(e,j|A,B)
    &=
    -\frac{1}{4\pi^2}
    \sum_{s=1}^{\infty}
    K(j,A-B;s)\,s^{-3}
    \left.
    \partial_{\mathcal B}^2
    R_{0,0}^{(1)}(e,j|\mathcal A,\mathcal B)
    \right|_{\mathcal A=A/s^2,\,\mathcal B=B/s^2}.
\end{split}
\label{eq:g2-R10-R01-perturbative}
\end{align}
The derivatives act only on the primitive contour kernel, not on the Kloosterman phase.  At the next mixed order,
\begin{align}
\begin{split}
    \rho_{1,1}(e_1,j_1;e_2,j_2)
    &=
    \mathcal N_1^2
    \sum_{\alpha,\beta\in\mathfrak P_2}
    C_\alpha\bar C_\beta
    \sum_{u_1,u_2,v_1,v_2\ge0}
    \prod_{i=1}^2
    c_{s_{\alpha i}}(u_i)c_{s_{\beta i}}(v_i)\\
    &\quad\times
    \prod_{i=1}^2
    R_{1,1}
    \left(
    e_i,j_i\Big|
    A_0+r_{\alpha i}+u_i,\,
    A_0+r_{\beta i}+v_i
    \right).
\end{split}
\label{5.32jj}
\end{align}
This term enters at order \(\epsilon^2\bar\epsilon^2\); at fourth order it must be combined with the \(F_2\overline F_0\) and \(F_0\overline F_2\) sectors. The two coefficients $\rho_{1,0}$ and $\rho_{0,1}$ are the first corrections in the holomorphic and anti-holomorphic sewing parameters, respectively.  They need not agree at fixed spin because they differentiate opposite energies.  Their exchange under simultaneous spin reflection is therefore a structural consequence of the computation, rather than an additional assumption about the spectrum.

\subsection{Numerical evaluation of handlebody OPE density }

\begin{figure}
    \centering
\includegraphics[width=0.48\linewidth]{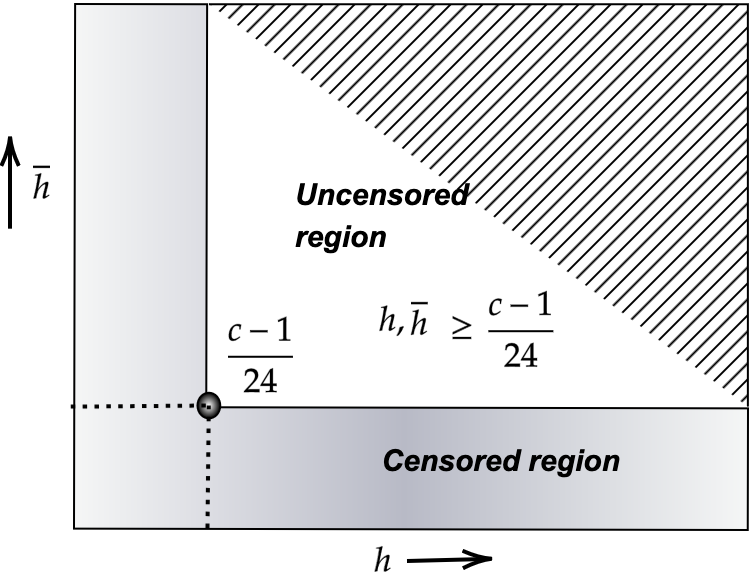}\hspace{1.5 cm}
    \caption{Figure depicting the light and heavy states (above the black hole threshold). The censored region contains only light operators. The heavy operator OPE density shows Cardy scaling. }
    \label{fig:censored-region}
\end{figure}

One of the primary goal of the article is to see whether the higher genus corrections to the density can cure the negativity of the spectrum near the black hole threshold. Our finding suggests that near the black hole threshold there exists a small region where the negativity still exists for the corrections $\rho_{1,0},\rho_{0,1}$. Negativity of $\rho_{0,1}$ appears for the one of the odd spin sector and the energy is chosen in the vicinity of the $|j|$ as depicted in Fig.~\eqref{fig:censored-region}.

To resolve the sign of the individual sewing coefficients, we parameterize
the energies by their distance from the BTZ threshold,
\begin{align}
    \delta_i=e_i-|j_i|>0,
    \qquad i=1,2.
\end{align}
The results below were obtained for $k=2$ using the kernel,
polar seed support, \textcolor{black}{Kloosterman cutoff} $s_{\max}=20$, Eisenstein cutoff $N_{E_2}=20$, and $80$-digit working precision. In the present implementation $s_{\max}$ is the cutoff on the Kloosterman sum.
Moreover, for $k=2$, the Eisenstein-mode sums cannot, in general, be initiated from the prescribed lower limit. Their lower limits must instead be adjusted according to the powers appearing in each term so that no spurious imaginary contributions arise in the density coefficients.
\begin{figure}[htb!]
    \centering
   \hspace{2 cm}\includegraphics[width=0.8\linewidth]{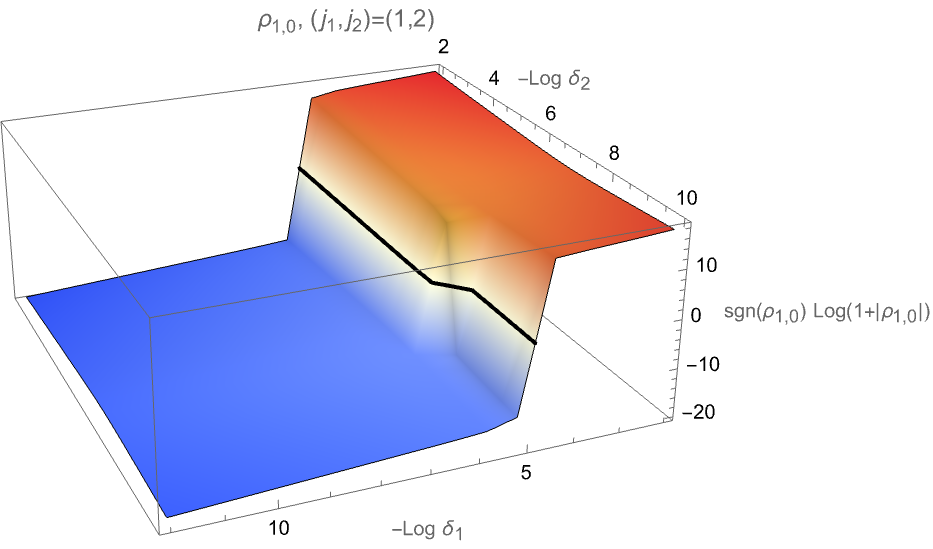}
    \caption{Plot depicting the nature of OPE density coefficient $\rho_{1,0}$ for different choices of $\delta_i$ with fixed $j_1=1, j_2=2$. We plot the signed logarithm $z\equiv \operatorname{sgn}(\rho_{1,0})\log\left(1+\lvert\rho_{1,0}\rvert\right)$. Thus, $z>0$ corresponds to $\rho_{1,0}>0$, $z<0$ corresponds to $\rho_{1,0}<0$, and $z=0$ corresponds to $\rho_{1,0}=0$. The orange region represents positive density, the blue region represents negative density, and the pale-yellow band indicates the transition between regions of large positive and large negative density. The black line marks the zero-density boundary. }
    \label{fid}
\end{figure}Alternatively, one may retain the original summation ranges and construct suitable combinations of the density coefficients using complex conjugation and spin reflection. These combinations yield a physically sensible real-valued density. We find numerically that the two prescriptions lead to the same qualitative conclusions, particularly regarding the occurrence and location of negative-density regions. We also check the convergence of the series by taking higher cut-off value and observe that the series converges very fast. Negative entries are shaded in red.
\begin{table}[]
    \centering
    \resizebox{\textwidth}{!}{%
    \begin{tabular}{rrrrrrrr}
        \toprule
        $j_1$ & $j_2$ & $\delta_1$ & $\delta_2$
        & $\rho_{0,0}$ & $\rho_{1,0}$ & $\rho_{0,1}$
        & $\rho_{1,1}^{(F_1\overline F_1)}$ \\
        \midrule
        $1$ & $1$ & $10^{-10}$ & $10^{-10}$
        & $ 2.79762\!\times\!10^{17}$ & $ 3.49089\!\times\!10^{16}$
        & $ 2.58039\!\times\!10^{-12}$ & $ 1.08130\!\times\!10^{-13}$ \\
        $1$ & $1$ & $10^{-8}$ & $10^{-4}$
        & \cellcolor{red!12}$-6.09309\!\times\!10^{14}$
        & \cellcolor{red!12}$-3.20933\!\times\!10^{13}$
        & $ 2.58333$ & $ 1.08259\!\times\!10^{-1}$ \\
        $1$ & $1$ & $10^{-6}$ & $10^{-2}$
        & \cellcolor{red!12}$-6.17194\!\times\!10^{14}$
        & \cellcolor{red!12}$-3.72042\!\times\!10^{13}$
        & $ 2.88920\!\times\!10^{6}$ & $ 1.21747\!\times\!10^{5}$ \\
        $1$ & $2$ & $10^{-8}$ & $10^{-2}$
        & \cellcolor{red!12}$-8.16016\!\times\!10^{18}$
        & \cellcolor{red!12}$-1.01989\!\times\!10^{18}$
        & $ 2.95532\!\times\!10^{6}$ & $ 2.60188\!\times\!10^{5}$ \\
        $1$ & $2$ & $10^{-6}$ & $10^{-2}$
        & \cellcolor{red!12}$-6.31602\!\times\!10^{17}$
        & \cellcolor{red!12}$-8.57537\!\times\!10^{16}$
        & $ 2.95535\!\times\!10^{9}$ & $ 2.60191\!\times\!10^{8}$ \\
        $1$ & $2$ & $10^{-4}$ & $10^{-2}$
        & $ 1.78529\!\times\!10^{18}$ & $ 1.54166\!\times\!10^{17}$
        & $ 2.95869\!\times\!10^{12}$ & $ 2.60499\!\times\!10^{11}$ \\
        $2$ & $2$ & $10^{-8}$ & $10^{-2}$
        & $ 2.17704\!\times\!10^{20}$ & $ 4.49519\!\times\!10^{19}$
        & $ 3.05842\!\times\!10^{9}$ & $ 5.67794\!\times\!10^{8}$ \\
        $3$ & $2$ & $10^{-10}$ & $10^{-2}$
        & \cellcolor{red!12}$-2.56275\!\times\!10^{22}$
        & \cellcolor{red!12}$-7.51681\!\times\!10^{21}$
        & $ 6.68433\!\times\!10^{8}$ & $ 1.88245\!\times\!10^{8}$ \\
        $3$ & $2$ & $10^{-8}$ & $10^{-2}$
        & $ 1.60861\!\times\!10^{21}$ & $ 4.23067\!\times\!10^{20}$
        & $ 6.68433\!\times\!10^{11}$ & $ 1.88246\!\times\!10^{11}$ \\
        $5$ & $2$ & $10^{-12}$ & $10^{-2}$
        & \cellcolor{red!12}$-1.67204\!\times\!10^{25}$
        & \cellcolor{red!12}$-7.86952\!\times\!10^{24}$
        & $ 3.66004\!\times\!10^{9}$ & $ 1.73238\!\times\!10^{9}$ \\
        $5$ & $2$ & $10^{-10}$ & $10^{-2}$
        & $ 6.12011\!\times\!10^{23}$ & $ 2.94144\!\times\!10^{23}$
        & $ 3.66004\!\times\!10^{12}$ & $ 1.73238\!\times\!10^{12}$ \\
        \bottomrule
    \end{tabular}}
    \caption{Table showing data for positive spin density coefficient
    The last column is the  $F_1\overline F_1$ contribution, $\rho_{1,1}$, to the second order correction, rather than the full fourth-order coefficient.}
    \label{tab1}
\end{table}
\begin{table}[]
    \centering
    \resizebox{\textwidth}{!}{%
    \begin{tabular}{rrrrrrrr}
        \toprule
        $j_1$ & $j_2$ & $\delta_1$ & $\delta_2$
        & $\rho_{0,0}$ & $\rho_{1,0}$ & $\rho_{0,1}$
        & $\rho_{1,1}^{(F_1\overline F_1)}$ \\
        \midrule
        $-1$ & $-2$ & $10^{-8}$ & $10^{-2}$
        & \cellcolor{red!12}$-8.16016\!\times\!10^{18}$
        & $ 2.95532\!\times\!10^{6}$
        & \cellcolor{red!12}$-1.01989\!\times\!10^{18}$
        & $ 2.60188\!\times\!10^{5}$ \\
        $-3$ & $-2$ & $10^{-10}$ & $10^{-2}$
        & \cellcolor{red!12}$-2.56275\!\times\!10^{22}$
        & $ 6.68433\!\times\!10^{8}$
        & \cellcolor{red!12}$-7.51681\!\times\!10^{21}$
        & $ 1.88245\!\times\!10^{8}$ \\
        $-5$ & $-2$ & $10^{-12}$ & $10^{-2}$
        & \cellcolor{red!12}$-1.67204\!\times\!10^{25}$
        & $ 3.66004\!\times\!10^{9}$
        & \cellcolor{red!12}$-7.86952\!\times\!10^{24}$
        & $ 1.73238\!\times\!10^{9}$ \\
        $1$ & $-2$ & $10^{-8}$ & $10^{-2}$
        & \cellcolor{red!12}$-8.16016\!\times\!10^{18}$
        & $ 2.85369\!\times\!10^{16}$
        & \cellcolor{red!12}$-3.95553\!\times\!10^{7}$
        & $ 2.76361\!\times\!10^{4}$ \\
        $-1$ & $2$ & $10^{-8}$ & $10^{-2}$
        & \cellcolor{red!12}$-8.16016\!\times\!10^{18}$
        & \cellcolor{red!12}$-3.95553\!\times\!10^{7}$
        & $ 2.85369\!\times\!10^{16}$
        & $ 2.76361\!\times\!10^{4}$ \\
        $2$ & $-2$ & $10^{-10}$ & $10^{-2}$
        & $ 1.98618\!\times\!10^{21}$
        & \cellcolor{red!12}$-7.55545\!\times\!10^{18}$
        & \cellcolor{red!12}$-4.09348\!\times\!10^{7}$
        & $ 7.57555\!\times\!10^{4}$ \\
        $-2$ & $2$ & $10^{-10}$ & $10^{-2}$
        & $ 1.98618\!\times\!10^{21}$
        & \cellcolor{red!12}$-4.09348\!\times\!10^{7}$
        & \cellcolor{red!12}$-7.55545\!\times\!10^{18}$
        & $ 7.57555\!\times\!10^{4}$ \\
        $2$ & $3$ & $10^{-10}$ & $10^{-2}$
        & $ 4.31797\!\times\!10^{23}$ & $ 1.36779\!\times\!10^{23}$
        & $ 6.64910\!\times\!10^{8}$ & $ 1.87087\!\times\!10^{8}$ \\
        $-2$ & $-3$ & $10^{-10}$ & $10^{-2}$
        & $ 4.31797\!\times\!10^{23}$ & $ 6.64910\!\times\!10^{8}$
        & $ 1.36779\!\times\!10^{23}$ & $ 1.87087\!\times\!10^{8}$ \\
        \bottomrule
    \end{tabular}}
    \caption{Table showing density coefficient data for reflected and mixed spin configuration.  Simultaneous spin reflection exchanges $\rho_{1,0}$ and
    $\rho_{0,1}$ while leaving the partial $\rho_{1,1}$ invariant.}
    \label{tab:genus2-high-cutoff-reflected}
\end{table}From the numerical analysis, we comes up with following observations. First, the negative bands move closer to the black hole threshold as the positive odd spin is increased.  At fixed
$\delta_2=10^{-2}$, the $(j_1,j_2)=(1,2)$ sector changes sign between $\delta_1=10^{-6}$ and $10^{-4}$, the $(3,2)$ sector between $10^{-10}$ and
$10^{-8}$, and the $(5,2)$ sector between $10^{-12}$ and $10^{-10}$.
Second, simultaneous spin reflection implements as follows,
\begin{align}
    \rho_{1,0}(j_1,j_2)
    &=\rho_{0,1}(-j_1,-j_2),
    &
    \rho_{1,1}^{(F_1\overline F_1)}(j_1,j_2)
    &=\rho_{1,1}^{(F_1\overline F_1)}(-j_1,-j_2).
\end{align}
\begin{figure}[b!]
    \centering
    \includegraphics[width=0.5\linewidth]{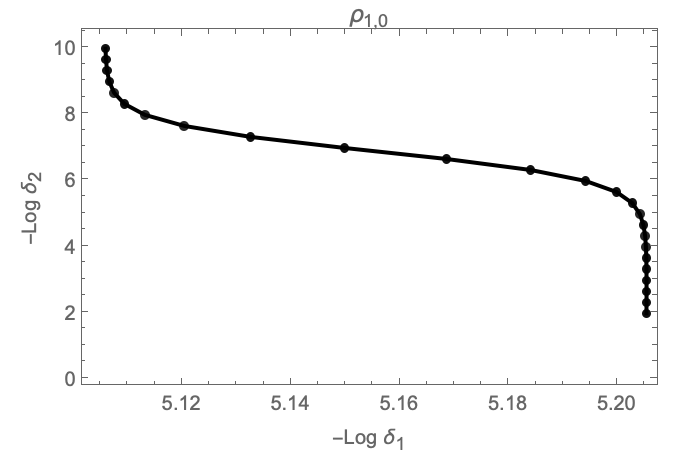}
    \caption{Figure depicting the zero density boundary for fixed $(j_1=1,j_2=2)$.}
    \label{zero}
\end{figure}
Third, mixed orientations can make both first-order chiral coefficients
negative, as occurs for $(j_1,j_2)=(2,-2)$ and $(-2,2)$ in
Table~\eqref{tab:genus2-high-cutoff-reflected}.  By contrast, the 
$F_1\overline F_1$ contribution to $\rho_{1,1}$ is positive at every point
listed here.  This last observation is a numerical statement on the scanned
domain, not a positivity theorem.  In particular, the complete fourth-order
coefficient also requires the presently unavailable
$F_2\overline F_0$ and $F_0\overline F_2$ contributions. In Fig.~\eqref{fid}, we provide a quantitative plot for the positive-negative transition of density coefficient $\rho_{1,0}$. As noted above, the zero-density condition defines a locus in the $(-\log\delta_1,-\log\delta_2)$ plane. In Fig.~\eqref{zero}, we display the numerically determined boundary separating the regions of large positive and large negative density.\\ \par

\noindent
\textbf{Stability of zero-density curve }
Since the truncated Kloosterman sum contains oscillatory contributions, it is natural to ask whether the zero-density locus remains stable as the cutoff is increased. Establishing this stability is important for determining whether the observed sign change is a genuine feature of the density rather than an artifact of truncating the sum. To test this, we consider the representative choice
$j_1=1$, $j_2=2$, and fix $\delta_2=10^{-2}$. For each value of the Kloosterman cutoff $s_{\max}$, we numerically determine the critical value $\delta_{1,\mathrm{crit}}$ satisfying
$$
\rho_{1,0}
\bigl(\delta_{1,\mathrm{crit}},\delta_2;
j_1,j_2\bigr)=0.
$$

\begin{figure}[t!]
    \centering
    \includegraphics[width=0.5\linewidth]{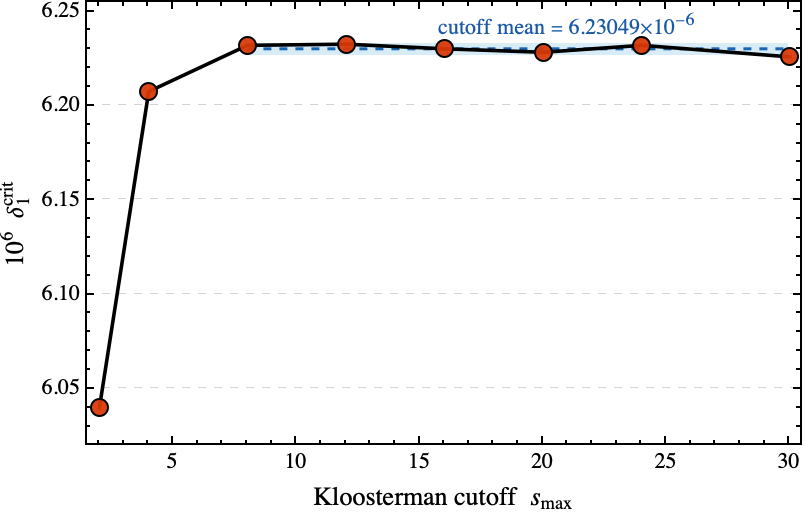}
    \caption{Dependence of the critical value
    $\delta_{1,\mathrm{crit}}$ on the Kloosterman cutoff $s_{\max}$
    for $j_1=1$, $j_2=2$, and $\delta_2=10^{-2}$.
    The dashed horizontal line denotes the mean obtained from the
    high-cutoff values $s_{\max}\in\{8,12,16,20,24,30\}$, while the
    shaded band indicates their minimum-to-maximum variation.}
    \label{fig:stab}
\end{figure}

As shown in Fig.~\eqref{fig:stab}, the critical value approaches a stable
plateau as $s_{\max}$ is increased. Averaging over the high-cutoff
values gives
\begin{equation}
    \overline{\delta}_{1,\mathrm{crit}}
    =
    \frac{1}{6}
    \sum_{s_{\max}\in\{8,12,16,20,24,30\}}
    \delta_{1,\mathrm{crit}}(s_{\max})
    =
    6.23049\times10^{-6}.
    \label{eq:mean-critical-delta}
\end{equation}
Over the same set of cutoffs, the critical values lie within
$$
6.22620\times10^{-6}
\leq
\delta_{1,\mathrm{crit}}
\leq
6.23294\times10^{-6}.
$$
The total variation is approximately $0.11\%$ of the mean value.
We therefore conclude that the zero-density crossing is stable under
increasing $s_{\max}$ and is not produced by a low-cutoff truncation.
\section{Conclusion and future outlooks}\label{sec5}
Motivated by the case study of calculating the OPE density above the black hole threshold at genus-one, we try to find out for higher genus whether the problem of non-unitarity of the spectrum has already been resolved or still remains.  In the following, we list the main findings of our paper.
\begin{itemize}
    \item We introspect the description of extremal conformal field theories on $g=2$ riemann surfaces. Using the techniques of sewing two riemann surfaces we figured out using modular invariance, how does the light spectrum (above the black hole threshold) looks like in gravity. The problem of negative density of states is not there for genus-two. Using the Rademacher methods we managed to find the exact form of the spectral density above the black hole threshold. We also showed the cardy growth of the spectrum for heavy primaries at genus-two.
\item 
We note that, in the \emph{separating degeneration limit}, the genus-two geometry effectively factorizes into the product of two genus-one topologies connected by a long tube. At least for splitted topologies with lower genus, connected by a thin long tube is easier for computing the genus-two OPE density rather than exploiting the full modular invariance of the dumbell and sunrise channel equality via crossing equation.

In particular, we show that the \emph{OPE density} associated with the genus-two handlebody admits a decomposition into a suitable combination of genus-one densities. Mathematically, at the next order in the pinching parameter, the genus-two contribution does not factorise into the products of genus-one data, and instead becomes two genus-one surfaces connected by a thin tube with the gluing parameter controlling the interaction between the two sectors.

Thus, in the separating limit, the complicated genus-two OPE density is described by simpler genus-one building blocks, providing a description of the handlebody contribution. Though, there are many non-handlebody instanton configuration is possible at genus-two, it is certainly important to exploit the total modular invariance and use modular bootstrap techniques to find the gravity OPE density similar to \cite{Simmons-Duffin:2025qox}. \textcolor{black}{We also show explicitly in the numerics that for very fine bands above the black hole threshold, we get the negativity. The negativity in genus-one density of state hence also persists in genus-two. This pathology of negativity, is not cured in genus-two. However, to strongly comment on this one needs to perform dumbell-sunrise channel bootstrap non-perturbatively. \textit{Apart from this we also showed explicitly, that the Rademacher contour prescription described in detail in appendix~\eqref{appA} renders different density than the obtained density from the direct inversion using the inverse laplace transform}. }

\noindent
\textbf{Future Outlooks.}\,\,Below we state some important future directions as follows,
\item\textit{\textbf{Spectral analysis:}} An immediate question concerns the issues of non-unitarity and discreteness of the spectrum. These problems are not yet resolved, and consequently, multiple possible resolutions remain open. In close analogy with the genus-one case, one option is to incorporate orbifold singularities. Another natural extension is to include non-handlebody instantons at genus two. From this perspective, it is particularly compelling to analyze the spectrum extracted from the partition function by exploiting modular invariance in the degeneration limit.
\textit{A further avenue of interest is the spectral analysis of the genus-two partition function itself, with particular emphasis on establishing bounds on the spectral gap associated with the lowest primary operators}.
\item \textit{\textbf{Functional bootstrap:} } It is important to investigate genus-two density of state using functional bootstrap, i.e writing the bootstrap equation on sphere with twist operator insertions, resembling the different higher genus structure.
\item\textit{\textbf{Matrix model description:}} An important future direction is to reproduce the spectral density obtained from the gravitational analysis directly from the proposed matrix model for pure three-dimensional gravity. In particular, one should verify that the matrix model topological recursion has a spectral curve which reproduces non-planar higher loop correlators, as the genus-two partition function as well as the density of states. Establishing this agreement would provide a novel consistency check of the matrix model duality and will clarify how much of the matrix model descriptions knew about the bulk topologies. 

\item \textit{\textbf{Performing the $Sp(4,\mathbb{Z})$ sum:}}A crucial aspect of a consistent genus-two formulation for extracting the OPE density is the implementation of full modular invariance under the total modular group $Sp(4,\mathbb{Z})$. In many approaches, one cosiders the diagonal subgroup $PSL(2,\mathbb{Z})\subset Sp(4,\mathbb{Z})$, corresponding to the separating degeneration limit where the genus-two surface factorizes into two tori.

This suggests that physical quantities can be expressed in terms of  Siegel modular forms and written as sums over the full $Sp(4,\mathbb{Z})$ modular group. Such a formulation would naturally incorporate both separating and non-separating contributions, avoiding reliance on degeneration limits \textit{Developing computational methods directly on the Siegel upper half-space, without invoking factorization, is therefore an important direction toward uncovering the complete modular structure.} 
\item \textit{\textbf{Non-Handlebody instantons}}: In this paper, we have computed the density of states arising from handlebody geometries, which constitute a seperate class of saddles admitting a simple geometric description in terms of contractible cycles. While these configurations capture an important subset of contributions, they are still subleading.
\end{itemize}

\section*{Acknowledgements}
 We thank Arpan Bhattacharyya for many useful discussions throughout the course of the project. S.P. (PMRF ID: 1703278) is supported by the Prime Minister’s Research Fellowship of the Government of India. Research of S.G. is supported by ANRF grant ANRF/ARG/2025/001338/PS. We would like to thank ChatGPT 5.6 and Codex for enormous help in writing the Mathematica code and formatting the draft.
\appendix
\section{Rademacher expansion of (holomorphic) modular integrals: a general strategy}\label{appA}
Let us assume a weakly holomorphic modular form of $SL(2,\mathbb{Z})$ with weight $w$,
\begin{align}
    \begin{split}
        f(\gamma\cdot \tau)=g(\gamma)(c\tau+d)^w f(\tau)
    \end{split}
\end{align}
\begin{figure}[htb!]
    \centering
\includegraphics[width=0.43\linewidth]{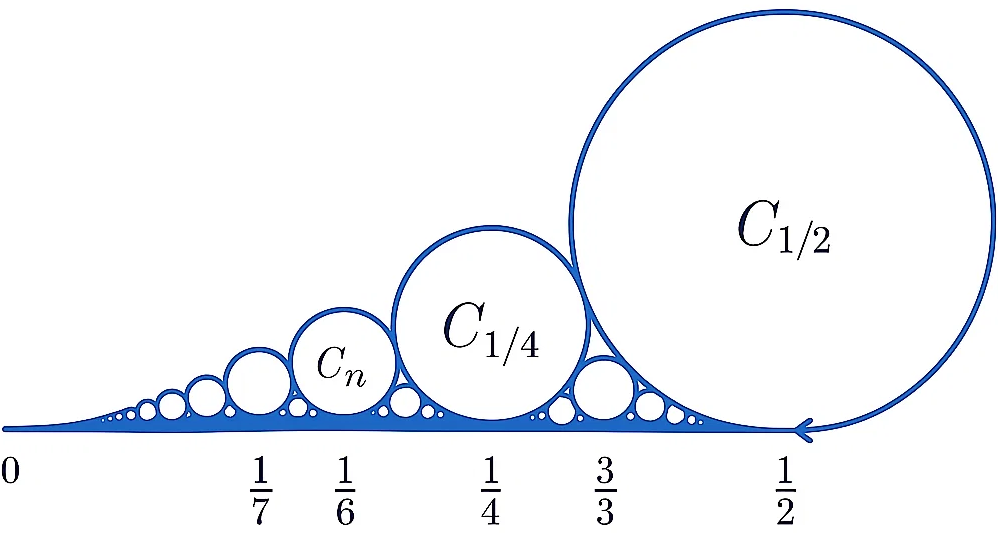}
    \caption{Ford circles obtained from farey sequences with sigularity at every $\frac{r}{s}$,where $(r,s)$ are co-prime to each other within the interval $(0,\frac{1}{2})$}
    \label{fig:9}
\end{figure}
\begin{figure}[htb!]
    \centering
\includegraphics[width=0.4\linewidth]{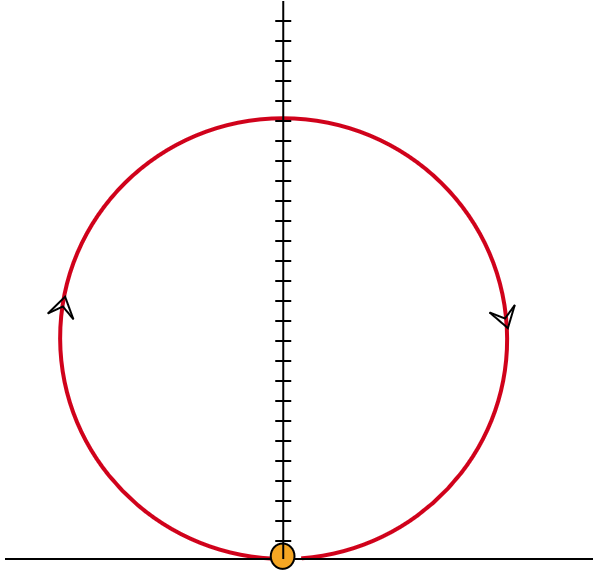}\hspace{1.5 cm}
\includegraphics[width=0.33\linewidth]{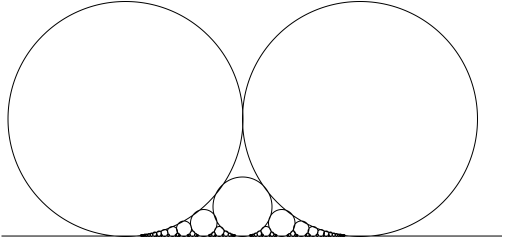}
    \caption{Here in the left figure we show the effective Rademacher contour $C_{01}$ which comes from each of the ford circles shown in the right. The ford cirlces are obtained from the \textit{Farey sequence. } }
    \label{fig:10}
\end{figure}
Assume, the modular form has polar part  and correspondingly has the following Fourier decomposition,
\begin{align}
    f(\tau)=\sum_{\mu=1}^{M}p_{-\mu}q^{-\mu}+\sum_{n\ge0}p_{n} q^n
\end{align}
\textit{{\bf Main goal}}: The main goal of (Hardy-Ramanujan-) Rademacher expansion is determine all the Fourier coefficient for $n\ge 0$ from the polar part of the expansion.
At first notice, the Fourier coefficients can be extracted straightforwardly from the Cauchy integral formula,
\begin{itemize}
    \item In $\tau$ variable the integral takes the form,
    \begin{align}
        p_{n}=\rmint_{\Gamma} f(\tau) e^{-2\pi i \tau n}\,d\tau,\,\tau=x+iy
    \end{align}
    \end{itemize}
Now assuming the function is holomorphic in the upper half plane, the contour in $q$-plane maps to $x\in [0,1]$ in $\tau$-plane and we choose the imaginary part to be 1 (it does not matter what we choose, confirmed from the holomorphicity). Therefore  the integral becomes,
\begin{align}
    p_{n}=\rmint_i^{i+1}d\tau\, f(\tau)\,e^{-2\pi i n \tau}.\label{A.5}
\end{align}
However a straightforward integral gives divergences. So the question is how one can find the convergent piece out of it. Rademacher's idea was to deform the contour in such a way that the resulting integral gives a convergent result, all thanks to the holomorphic (and periodicity) nature of the modular function.\\

\noindent
 \textit{{\bf Farey dissection and Rademacharisation}}: Enlarge the contour until it touches the real line. The contour deforms into a union of Ford-circle arcs $\Gamma_{r/s}$ around each reduced rationals $r/s$ (Farey dissection) with $s\le N$. All the Ford circles centers at $\tau=\frac{r}{s}+i\frac{1}{2s^2}$ with radius $\frac{1}{2s^2}$ as desribed in Fig.~\eqref{fig:9} and Fig.~\eqref{fig:10}. Effectively, we replace the straight line contour with union of the arcs of the Ford circles. Therefore the Fourier coefficients become,
\begin{align}
    p_{n}=\sum_{\substack{r/s \,\in\, \mathcal{F}_N \\[1pt] 0 < r \le s, (r,s)=1 \\[1pt] s\le N}}
\;\rmint_{\Gamma_{r/s}}
f(\tau)\,e^{-2\pi i n \tau}\,d\tau
\end{align}
 Now straighten the each arc by the following modular map,
\begin{align}
    \gamma_{r/s}=\begin{pmatrix}a&b\\ s&-r\end{pmatrix}\in\mathrm{SL}(2,\mathbb Z), \,\textrm{gcd}(r,s)=1
\end{align}
after using the modular transformation properties the integral becomes,
\begin{align}
    \begin{split}
    \rmint f(\tau)e^{-2\pi i \tau n}&\to \sum_{s=1}^{N}\sum_{\substack{ 0 < r \le s,\\[1pt] (r,s)=1 }}g^{-1}(\gamma)\rmint_{\Gamma_{r/s}} d\tau \,e^{-2\pi  in \tau} (s\tau-r)^{-w}f\left(\frac{a\tau+b}{s\tau -r}\right) \\ &
    =\sum_{s=1}^{N}\sum_{\substack{ 0 < r \le s,\\[1pt] (r,s)=1 }}\sum_{\mu=1}^{M} p_{-\mu}  g^{-1}(\gamma)\rmint d\tau \,(s\tau-r)^{-w}\\&\hspace{3 cm}\times\exp\left(-2\pi i \mu\left(\frac{a\tau+b}{s\tau-r}\right)-2\pi i n \tau \right)\\ &
    =\sum_{s=1}^{N}\sum_{\substack{ 0 < r \le s,\\[1pt] (r,s)=1 }}\sum_{\mu=1}^{M} p_{-\mu}  g^{-1}(\gamma)\rmint d\tau \,(s\tau-r)^{-w}\\&\hspace{3 cm}\times\exp\left(-2\pi i \mu\left(\frac{a}{s}-\frac{1}{s(s\tau-r)}\right)-2\pi i n \tau \right)
    \end{split}
\end{align}
Now, do the following change of variable, $s\tau-r=\frac{iz}{s}$ (which sends every Ford circle to the origin with radius one.), we have ,
\begin{align}
    \begin{split}
        i^{1-w}\sum_{s=1}^{N}s^{w-2}\sum_{\substack{ 0 < r \le s,\\[1pt] (r,s)=1 }}\sum_{\mu=1}^{M} p_{-\mu}  g^{-1}(\gamma)\rmint_{z_1(r,s,N)}^{z_2(r,s,N)} dz\,z^{-w}\,e^{-\frac{2\pi i}{s}(nr+\mu a)}\,\exp\left[\frac{2\pi \mu }{z}+\frac{2\pi n z}{s^2}\right]
    \end{split}
\end{align}
As we take, $N\to \infty$, the Rademacher contour almost encircles the full Ford circles except the point $z=0$ (in $\tau$-plane $\tau=r/s$). Now again doing the following change of variable which maps the circle to a vertical straight line with $\Re(t)=2\pi$, $z=\frac{2\pi}{t}$, we have,
\begin{align}
    \begin{split}
&    -   (2\pi i)^{ 
       1-w}  \sum_{s=1}^{N}s^{w-2}\sum_{\mu=1}^{M} p_{-\mu}  \left(\sum_{\substack{ 0 < r \le s,\\[1pt] (r,s)=1 }} g^{-1}(\gamma)\,e^{-\frac{2\pi i}{s}(nr+\mu a)} \right)\rmint_{2\pi-i\infty}^{2\pi+i \infty} dt\,t^{w-2} e^{\mu t+\frac{4\pi^2 n}{t s^2}}\\ &
       =-(2\pi i)^{2-w}\sum_{\mu=1}^{M} p_{-\mu}\,   \mu^{1-w} \,\sum_{s=1}^{N\to \infty}\frac{1}{s}K_{g}(n,
    \mu,s) \left({2\pi \sqrt{\mu n}}\right)^{w-1} I_{1-w}\left(\frac{4\pi \sqrt{\mu n}}{s}\right),
    \end{split}
\end{align}
where, $K_{g}(\cdots)$ is the generalized Kloosterman sum defined as,
\begin{align}
    K_{g}(n,\mu,s)=\sum_{\substack{ 0 < r \le s,\\[1pt] (r,s)=1 }} g^{-1}(\gamma)\,e^{-\frac{2\pi i}{s}(nr+\mu a)}
\end{align}

\subsection{ Rademacher contour for OPE density in CFT}\label{4.1w}
In this section we describe how to compute the density of states using the Rademacher contour deformation. For a detailed description of the method we refer the reader to appendix~\eqref{appA}. The question we want to ask is the following: for a given isolated operator ($h_0,\bar h_0$), such that $h_0+\bar h_0<2\hat c$\footnote{with $\hat c=\frac{c-1}{24}$}, what would be the implications of the density $d_{h\bar h}$ on the left side?\\

Let's first focus on the LHS of the bootstrap equation. 
\begin{align}
    \begin{split}
        \sum_{h,\bar h}d_{h\bar h} e^{2\pi i \tau (h-\hat c)} e^{-2\pi i \bar\tau(\bar h-\hat c)}\xrightarrow[\tau\to x+i y]{\bar{\tau}\to x-iy}\sum_{j,\Delta}d_{j\Delta}\, e^{2\pi i jx}e^{-2\pi y \, e}, \,\textrm{with,}\, j=h-\bar h,
    \end{split}
\end{align}
Formally, the OPE density can be found as,
\begin{align}
\begin{split}
  &  \sum_{\Delta}\frac{d_{j\Delta}}{2\pi i}\rmint_{\varepsilon-i\infty}^{\varepsilon+i \infty} e^{-2\pi y e+2\pi y e'}dy=\frac{1}{2\pi i}\rmint_{\varepsilon-i\infty}^{\varepsilon+i \infty}a_{j}(y)e^{2\pi y e'} dy\\ &
  \implies \sum_{\Delta}{d_{je}}\,\delta(e-e')=\frac{1}{2\pi i}\rmint_{\varepsilon-i\infty}^{\varepsilon+i \infty}a_{j}(y)e^{2\pi y e'} dy\\ &
  \implies d_{je}=\frac{1}{2\pi i}\rmint_{\varepsilon-i\infty}^{\varepsilon+i \infty}a_{j}(y)e^{2\pi y e} dy\label{4.39r}
  \end{split}
\end{align}

\noindent
There is a relation for each pair of positive coprimes $(r, s)$. Now parametrizing these in terms of spin and conformal dimensions of the operators in the spectrum we achieve (from the RHS),

\begin{align}
\begin{split}
    \mathbfcal{H}(x, y) &=  \left(\frac{y}{y'}\right)^b\, e^{2\pi i \left(\frac{a\tau+b}{s\tau-r}\right)(h_0-\hat c)}e^{-2\pi i \left(\frac{a\bar\tau+b}{s\bar\tau-r}\right)(h_0-\hat c)}\,\,\,\,\,\\ &
      =\left[(sx-r)^2+s^2y^2\right]^b \,e^{2\pi i \left(\frac{a}{s}-\frac{1}{s(s\tau-r)}\right)(h_0-\hat c)}e^{-2\pi i \left(\frac{a}{s}-\frac{1}{s(s\bar\tau-r)}\right)(\bar h_0-\hat c)}\\ &
      =\left[(sx-r)^2+s^2y^2\right]^b  e^{2\pi i \frac{a}{s}J}e^{-\frac{2\pi i }{s}\left(\frac{h_0-\hat c}{s\tau -r}-\frac{\bar{h}_0-\hat c}{s\bar \tau-r}\right)}
      \\ &
      =\left[(sx-r)^2+s^2y^2\right]^b  e^{2\pi i \frac{a}{s}J} e^{-\frac{2\pi i}{s}\frac{sx-r}{(sx-r)^2+s^2 y^2}J} e^{-\frac{2\pi y E}{(sx-r)^2+s^2 y^2}}\\&
= s^{2b} \left(\left(x-\frac{r}{s}\right)^2+y^2\right)^b 
\, e^{\frac{2\pi i a}{s} J} 
\, e^{\frac{2\pi i (r - s x)}{s \left((r - s x)^2 + s^2 y^2 \right)} J} 
\, e^{-\frac{2\pi y E}{(r - s x)^2 + s^2 y^2}}
\end{split}
\end{align}
where we used, \begin{align}
    y'=\frac{y}{(sx-r)^2+s^2 y^2},\,\,\,e=\Delta-2\hat c,\,\,\,\,E:=\Delta_0-2\hat c, \,\,\,\,J:=h_0-\bar h_0
\end{align}
alongwith $\Delta_0=h_0+\bar h_0$ and $\hat c=\frac{c-1}{24}$. The main goal of this section is to find the density of the light states which are conical defects and well below the  blackhole threshold exploiting the modular invariance. Now, we can recast the sum as,
\begin{align}
    \sum_{j=-\infty}^{\infty}a_j(y)e^{2\pi i j x}=\mathbfcal{H}(x, y) 
\end{align}
Now, using the fourier analysis and the modular invariance we can extract the OPE coefficient as (we extract the modular coefficient with a generic power $b\in \mathbb{Z}$
of the integrand i.e $(x^2+y^2)^b$),
\begin{align}
\begin{split}
    a_{j}(y):&=\rmint_0^1 \mathbfcal{H}(x,y)e^{-2\pi i j x} dx \end{split}
\end{align} 
As it can be seen, it has similar form of \eqref{A.5}, which can be evaluated using the technology of \textit{Rademacher contours as mentioned in the Appendix~\eqref{appA}, deforming the straightline contour into sum of ford circle} we get,
Hence, deforming the straightline contour into sum of ford circle we get,
\begin{align}
    \begin{split}
        \rmint_0^1 \mathcal{H}(x,y)\,e^{-2\pi i j x} dx &\to \sum_{s=1}^N \sum_{\substack{ 0 < r \le s,\\[1pt] (r,s)=1 }}\rmint_{\Gamma_{r/s}} dx e^{-2\pi i j x} \mathcal{H}(x,y)
    \end{split}
\end{align}
Now doing the following change of variable, $sx-r={is(z-y),}$ we get,
\begin{align}
   i \sum_{s=1}^N \sum_{\substack{ 0 < r \le s,\\[1pt] (r,s)=1 }}e^{2\pi i \frac{a}{s}J-2\pi i j \frac{r}{s}}\rmint_{z_1}^{z_2} dz\, (-1)^b\,s^{2b}(z^2-2yz)^{b} e^{-\frac{2\pi}{s^2}\frac{z-y}{z^2-2y z}J}\,e^{\frac{2\pi y E}{s^2(z^2-2y z)}} e^{2\pi  j (z-y)} 
\end{align}
Now unfolding the circle to a vertical straight line by the following transformation: $z=\frac{2\pi}{t}$ and sending $N\to \infty$ we get,
\begin{align}
    \begin{split}
      & \hspace{1 cm} -2\pi(-1)^b s^{2b}\rmint_{\sigma-i\infty}^{\sigma+i\infty} \frac{dt}{t^2}e^{2\pi j\left(\frac{2\pi }{t}-y\right)} \left(\frac{4\pi^2}{t^2}\right)^b \left(1-\frac{y\,t}{\pi}\right)^b \\&\hspace{4 cm}\times \exp\left(-\frac{2\pi}{s^2}\frac{\frac{2\pi}{t}-y}{\frac{4\pi^2}{t^2}\left(1-\frac{yt}{\pi}\right)}J\right) \exp\left(\frac{2\pi y}{s^2}\frac{\pi t^2}{4\pi^2 (\pi-yt)}E\right)\\ &
      \hspace{1 cm} = -2\pi(-1)^b s^{2b}\rmint_{\sigma-i\infty}^{\sigma+i\infty} \frac{dt}{t^2}e^{2\pi j\left(\frac{2\pi }{t}-y\right)} \left(\frac{4\pi^2}{t^2}\right)^b \left(1-\frac{y\,t}{\pi}\right)^b\\&\hspace{4 cm}\times 
      \exp\left(\frac{t}{2\pi }\frac{2\pi-yt}{\frac{yt }{\pi}-1}{\color{black}{\frac{\mathbf{J}}{\mathbf{s}^2}}}\right)\exp\left(-\frac{yt^2}{2 \pi \left(1-\frac{y t }{\pi}\right)}{\color{black}{\frac{\mathbf{E}}{\mathbf{s}^2}}}\right)
    \end{split}
\end{align}
Therefore the Fourier coefficient is given by,
\begin{align}
    \begin{split}
        a_{j}(y)&=\left(\frac{1}{4\pi^2}\right)^b \sum_{s=1}^{\infty}s^{2b}\,K(j,J,s)\times\\&\frac{1}{2\pi i}\rmint_{\sigma-i\infty}^{\sigma+i\infty} \frac{dw}{w^{2+2b}} e^{j\left(\frac{1}{w}-2\pi y\right)}(4\pi y w-1)^b\,\\&\hspace{2 cm}
    \times \exp\left(\frac{4\pi^2w (1-2\pi y w)}{4\pi y w-1}\frac{J}{s^2}\right)\exp\left(-\frac{8 y \pi^3 w^2}{4\pi y w-1}\frac{E}{s^2}\right)\\&
   =\left(\frac{1}{4\pi^2}\right)^b \sum_{s=1}^{\infty}s^{2b}\,K(j,J,s) d_{s}(y,E,J)
    \end{split}
\end{align}
It is instructive to analyze the large $\omega$ limit, more precisely, $\omega \kappa_{+} \gg 1$. In this limit, $d_{s}(y,E,J)$ becomes $(\text{\,\,\,with},
\,\,\,\,   {\kappa_{\pm}=-(E\pm J)})$,
\begin{align}
    \begin{split}
       & d_{s}(y,E,J)=\frac{1}{2\pi i}\rmint \frac{dw}{w^{2+2b}} \,e^{j\left(\frac{1}{w}-2\pi y\right)}(4\pi y w-1)^{b}\exp\left(\frac{2\pi^2w (\kappa_{-}-\kappa_{+})}{4\pi y w-1}\right)\exp\left(\frac{8y\pi^3 w^2}{4\pi yw-1}\kappa_{+}\right)\,\\&
       \xrightarrow[ ]{\omega \kappa_{+} \gg 1}e^{-\frac{\pi}{2y}\kappa_{+}}\frac{1}{2\pi i}\rmint \frac{dw}{w^{2+2b}} \,e^{j\left(\frac{1}{w}-2\pi y\right)}(4\pi y w-1)^{b}\exp\left(2\pi^2 w\frac{ \kappa_{+}}{s^2}\right) \exp\left(\frac{2\pi^2 w }{4\pi y w-1}\frac{\kappa_{-}}{s^2}\right)
    \end{split}\label{5.21j}
\end{align}
As evident from \eqref{5.21j}, $d^{s}(y,E,J)=d^{s=1}(y,E,J)\Big|_{E\to \frac{E}{s^2},\,J\to \frac{J}{s^2}}$.
Therefore, formally the OPE density is given by,
\begin{align}
    d(j,e)=\left(\frac{1}{4\pi^2}\right)^b\sum_{s=1}^{\infty}s^{2b}\,K(j,J,s)\, \mathcal{L}_{y\to e}^{-1}\,d_{s}(y,E,J)
\end{align}

\noindent
Now using \eqref{4.39r} we can explicitly compute $a_j(y)$ as follows,
\begin{align}
\begin{split}
    a_{j}(y):&=\rmint_{\frac{1}{2\pi}-i\infty} ^{\frac{1}{2\pi}+i\infty} 4^{-b} \pi ^{-2 b} \left(\frac{4 \pi  u y-1}{u^2}\right)^b \textstyle{\exp \left(\frac{4 \pi ^2 u^2 (-2 \pi  E u y-2 \pi  J u y+J)+j \left(-8 \pi ^2 u^2 y^2+6 \pi  u y-1\right)}{u (4 \pi  u y-1)}\right)\frac{-i}{ u^2}du}\\&
   =\rmint_{\frac{1}{2\pi}-i\infty} ^{\frac{1}{2\pi}+i\infty} 4^{-b} \pi ^{-2 b} \left(\frac{4 \pi  u y-1}{u^2}\right)^b e^{(\frac{1}{u}-2\pi y)j}
e^\frac{{-2\pi^2 u}\kappa_{-}}{1 - 4\pi u y }  e^{2\pi^2 \kappa_{+} u}\,\,\frac{-i}{ u^2}du , \,\texttt{\,\,\,with},
\,\,\,\,   {\kappa_{\pm}=-(E\pm J)}\\&
=\sum_{t,q}\rmint_{\frac{1}{2\pi}-i\infty} ^{\frac{1}{2\pi}+i\infty} \frac{4^{-b} \pi ^{-2 b} }{t!q!}\left(\frac{4 \pi  u y-1}{u^2}\right)^b e^{-2\pi yj}\Bigg(\frac{j}{u}\Bigg)^{t}
\Bigg(\frac{{-2\pi^2 u}\kappa_{-}}{1 - 4\pi u y }\Bigg)^q  e^{2\pi^2 \kappa_{+} u}\,\,\frac{-i}{ u^2}du\\&
\end{split}
\end{align}
In the last line we assumed $u$ to be large and approximated exponential appropriately.

\noindent
Now, using the following integral identity,
\begin{align}
    \begin{split}
\frac{1}{2\pi i}
\rmint_{\frac{1}{2\pi } - i\infty}^{\frac{1}{2\pi } + i\infty}
\frac{u^\alpha}{(4\pi u y - 1)^\beta} \, e^{2\pi^2 \kappa_+ u} \, du
=\frac{(\!-\kappa_+)^{-\alpha + \beta - 1}}{2^{\alpha + \beta + 1} \pi^{2\alpha - \beta + 2} y^\beta}\cdot \Gamma(\beta - \alpha) 
{}_1F_1\left( \beta;\, \beta - \alpha;\, \kappa_+ \frac{\pi}{2y} \right)
   \end{split}
\end{align}
Hence, the fourier coefficient is given by,
\begin{align}
\begin{split}
    a_j&(y)=\sum_{t,q}\frac{j^{t}e^{-2\pi y j}
(-2\pi^2\kappa_{-})^q }{q!\,t!}\frac{(\!-\kappa_+)^{-\alpha + \beta - 1}}{2^{\alpha + \beta + 1} \pi^{2\alpha - \beta + 2} y^\beta}
 \\&\hspace{3 cm}\times\Gamma(\beta - \alpha) 
{}_1F_1\left( \beta;\, \beta - \alpha;\, \kappa_+ \frac{\pi}{2y} \right)\Bigg|_{\alpha=q-t-2-2b,\beta=-b+q}\\&
\end{split}
\end{align}
Now, we can write the result of our density of state at $g=1$ by performing inverse Laplace transform\footnote{While performing the inverse Laplace transform we use the identity,
\begin{align}
    \mathcal{L}^{-1}\bigg(\frac{\exp (-2 \pi  j y)}{y^m}\bigg)=\frac{(2 \pi )^{m-1} \Theta (e-j) (e-j)^{m-1}}{\Gamma (m)}
\end{align}} as (for $b=-\frac{1}{2}$),
\begin{align}
\begin{split}&\rho^{s=1}_{-1/2;E,J}(e,j)=\frac{1}{{\sqrt{e^2-j^2}}}\bigg[\sinh \left(\frac{\sqrt{2} \pi  j \left(\sqrt{\kappa_{-}}+\sqrt{\kappa_{+}}\right)}{\eta }-\sqrt{2} \pi  \eta  \left(\sqrt{\kappa_{-}}-\sqrt{\kappa_{+}}\right)\right)\\&\hspace{5 cm}-\sinh \left(\frac{\sqrt{2} \pi  j \left(\sqrt{\kappa_{-}}-\sqrt{\kappa_{+}}\right)}{\eta }-\sqrt{2} \pi  \eta  \left(\sqrt{\kappa_{-}}+\sqrt{\kappa_{+}}\right)\right)\bigg]
\label{3.37r}
\end{split}
\end{align}
where, $\eta:= \sqrt{\sqrt{e^2-j^2}+e}$.
Now, this is very specific to $s=1$, further we need to sum over $r,s$ i.e the Rademacher parametrization\footnote{Now using the following identity,
\begin{align}
I_\nu(z) := \frac{( \tfrac{1}{2} z)^{\nu}}{\Gamma(\nu + 1)} \, {}_0F_1\left(-;\nu + 1; \tfrac{1}{4} z^2\right) 
= \sum_{n=0}^{\infty} \frac{(\tfrac{1}{2} z)^{\nu + 2n}}{\Gamma(\nu + n + 1) \, n!}
\end{align} one can also cast the result in terms of modified Bessel function.} and the sum is given by,
\begin{align}
\begin{split}
\rho_{b;E,J}(e,j):&=\sum_{s=1}^{N}\sum_{\substack{ 0 < r \le s,\\[1pt] (r,s)=1 }}
e^{2\pi i\left(-\frac{r}{s} j + \frac{a{[r,s]}}{s} J\right)} \, \rho^{s=1}_{(b;\frac{E}{s^2},\frac{J}{s^2})}(e,j)\,s^{2b}\\&
=\sum_{s=1}^\infty K(j,J,s)s^{2b} \,\times\rho^{s=1}_{(b;\frac{E}{s^2},\frac{J}{s^2})}(e,j)\label{eq:grav-rademacher-sum}
\end{split}
\end{align}


\noindent
The generic $s$ terms are obtained from the $s=1$ case via scaling the $(E,J):\to(\frac{E}{s^2},\frac{J}{s^2})$ defined before. Though we consider a single operator above the BTZ threshold one can generailze it to a density of operators or multiple operators to extract the OPE density of light operators. One can show that the sum is actually covergent and divergent both for different scenario(s). We cast it in Table~\eqref{Tab3}.
\noindent
We used the \textit{ M$\ddot{o}$bius $\mu$ function} and \textit{Euler totient $\phi$ function} for performing the sums.
From the Table~\eqref{Tab3}, it is evident that the scalar operator density (which we mainly focus on) below the black hole threshold is convergent only in certain cases. Maintaining a nonzero spin for operators above the threshold yields a more substantial, nonvanishing, and convergent OPE density. We checked numerically that, while the closed form is absent for \textit{the Kloosterman sum} for more higher spins ($J$), they are more convergent for scalar OPE density below the threshold. As we have the basic OPE density found from the modular invariance of CFT partition function, we now proceed to comment on the light state OPE density for gravity for genus two euclidean geometries.

\section{{Performing the modular sum}}
\label{appB}Following \cite{Maloney:2007ud}, we can rewrite the sum in \eqref{2.21e} as follows,
\begin{align}
\begin{split}
   \tilde{\mathbfcal{E} }(\kappa,\mu,s_1,w)&:=\sum_{c,d}|c\tau+d|^{{2s_1+2w-1}}\exp(2\pi \kappa \,\Im(\gamma\cdot\tau)+2\pi i \mu\, \Re(\gamma\cdot \tau)),\\&
    =e^{2\pi(\kappa y+i\mu x)}+\sum_{c>0}\sum_{d'\in \mathbb{Z}/c\mathbb{Z}}\sum_{n\in \mathbb{Z}}\mathcal{S}(c,d',n)\label{3.14r},
\end{split}
\end{align}
    where the first term comes from the $c=0,d=1$ and the remaining summand is given by,
    \begin{align}
        \mathcal{S}(c,d',n):&=\frac{1}{|c(\tau+n)+d'|^{2s_1+2w-1}}\,\\&\times\exp\Bigg(\frac{2\pi \kappa y}{|c(\tau+n)+d'|^{2}}+{2\pi i\mu}\Bigg(\frac{a}{c}{-\frac{cx+d}{c(|c(\tau+n)+d'|^2)}}\Bigg)\Bigg)
    \end{align}
Now the sum written in \eqref{3.14r} can be performed and is given by the following expression:\footnote{For details we refer the reader to \cite{Maloney:2007ud}.} 
\begin{align}
    \tilde{\mathbfcal{E} }(\kappa,\mu,\tilde s)&:=e^{2\pi(\kappa y+i\mu x)}+\sum_{\hat n}e^{2\pi i\hat n x}\mathbfcal{E}_{\hat n}(\tilde s,\kappa,\mu)
\end{align}
 with the following fourier summand,

 \begin{align}
     \mathbfcal{E}_{\hat n}(\tilde s,\kappa,\mu)=\sum_{p=0}^\infty \mathbfcal{I}_{p,\hat n}(\tilde s,\kappa,\mu)y^{1-p-\tilde s}\underbrace{\Bigg(\sum_{c=1}^\infty c^{-2(p+ \tilde s)}K(-\hat n,\mu;c)}_{\mathbfcal{T}}\Bigg),\label{3.8u}
 \end{align}
 where we consider,
\begin{align}
    \tilde s:=1-s_1-w
\end{align}

\noindent
alongwith, the \textit{Kloosterman sum} defined as,
\begin{align}
    K(-\hat n,\mu;c)=\sum_{d\in (\mathbb{Z}/c{\mathbb{Z})^*}}\exp\bigg[2\pi i \Bigg(\frac{-\hat n d+\mu d^{-1}}{c}\Bigg)\bigg].
\end{align}
We also have the integral present in the sum can be written as \cite{Maloney:2007ud},
\begin{align}
    \begin{split}
        \mathbfcal{I}_{p,\hat n}(\tilde s,\kappa,\mu):=\textcolor{black}{\frac{(2\pi)^p}{p!}}\rmint_{-\infty}^\infty dT e^{2\pi i \hat n T y}(1+T^2)^{-p-\tilde s}(\kappa-i\mu T)^p
    \end{split}
\end{align}
One should note that it is independent of $x$. Now we discuss a few cases,
\begin{itemize}
\item \textbf{ Case-1: $\hat n=0$ mode with\,$\mu=0$}
\begin{align}
    \mathbfcal{I}_{p,0}(\tilde s,\kappa,\mu=0)=\kappa^p\frac{2^p\pi^{p+1/2}\Gamma(\tilde s+p-\frac{1}{2})}{p!\,\Gamma(\tilde s+p)}
\end{align}
for this case, we can cast the sum over $c$ in the parantheses in \eqref{3.8u} as \cite{Maloney:2007ud},
\begin{align}
    \sum_{c=1}^{\infty}{c}^{-2+p+\tilde s}K(0,0;c)=\frac{\zeta(-(p+\tilde s)-3)}{\zeta(-2-(p+\tilde s))}
\end{align}

 \item \textbf{Case-2: $\hat n=0$ mode with\,$\mu=\pm1$}
\begin{align}
\begin{split} 
&\mathbfcal{I}_{p,0}(\tilde s, \kappa, \pm 1) =
\cos\!\left(\frac{p\pi}{2}\right) (2\pi)^{p} 
\frac{\Gamma\!\left( \frac{1+p}{2} \right) \Gamma\!\left( \frac{p-1}{2} +\tilde  s \right)}{p!\,\Gamma(p+\tilde s)}
\, {}_{2}F_{1}\!\left( \frac{p-1}{2} + \tilde s,\; -\frac{p}{2} +\tilde s;\; \frac{1}{2};\; \kappa^{2} \right)
\quad \\&+\; p\kappa \, \sin\!\left(\frac{p\pi}{2}\right) (2\pi)^{p} 
\frac{\Gamma\!\left( \frac{p}{2} \right) \Gamma\!\left( \frac{p}{2} + \tilde s \right)}{p!\,\Gamma(p+\tilde s)}
\, {}_{2}F_{1}\!\left( 1 - \frac{p}{2},\; \frac{p}{2} +\tilde s;\; \frac{3}{2};\; \kappa^{2} \right)
\end{split}
\end{align}
\item \textbf{Case-3: $\hat n=0$ mode with\,$\mu=\pm2$ }
\begin{align}
\begin{split} 
&\mathbfcal{I}_{p,0}(\tilde s, \kappa, \pm 2) =-\frac{\pi ^{3/2} \kappa ^p (2\pi)^p\sec (\pi  (p+\tilde s)) \, _2F_1\left(\frac{1-p}{2},-\frac{p}{2};-p-\tilde s+\frac{3}{2};\frac{1}{\kappa ^2}\right)}{p!\,\Gamma \left(-p-\tilde s+\frac{3}{2}\right) \Gamma (p+\tilde s)}\\&+\frac{i \pi(2\pi)^p  \left((-i \kappa )^{2 (p+\tilde s)}-(i \kappa )^{2 (p+\tilde s)}\right) \kappa ^{-3 p-4 \tilde s+1} \csc (2 \pi  (p+\tilde s)) \Gamma (p+2 \tilde s-1) \, }{p!\,\Gamma (-p) \Gamma (2 (p+ \tilde s))} \\&\hspace{4 cm}\times \,_2F_1\left(\frac{p-1}{2}+ \tilde s,\frac{p}{2}+\tilde s;p+ \tilde s+\frac{1}{2};\frac{1}{\kappa ^2}\right) \texttt{\,\,\,\,\,for $p \neq 0$}\\&
=\sqrt{\pi}\frac{\Gamma(\tilde s-1/2)}{\Gamma(\tilde s)}\,\,\,\,\,\,\, \texttt{for $p=0$},
\end{split}
\end{align}
\end{itemize}
for this case, we can cast the sum over $d$ in the parantheses in \eqref{3.8u} as
\begin{align}
\begin{split}
     S(0,\pm 2;c)&=\mu(c);\,\,\,\,\,\, c \in \texttt{Odd integers}\\&
=\mu(c)+2\mu(c/2);\,\,\,\,\,\, c \in \texttt{even integers},
\end{split}
\end{align}
where $\mu(c)$ is known as the M$\ddot{o}$bius function.
Now the whole sum over $c$ can be written as ,
\begin{align}
\begin{split}
\mathbfcal{T}&:=\sum_{c_{odd}=1}^\infty c^{-2(p+\tilde s)}\mu(c)+\sum_{c_{even}=2}^\infty c^{-2(p+\tilde s)}[\mu(c)+2\mu(c/2)]\\&=\frac{1}{\zeta(2(\tilde s+p))}\Bigg(1+2^{1-2p-2\tilde s}\Bigg)
\end{split}
\end{align}
\noindent
Therefore we get the following answer for the sum:

    \begin{align}
     \mathbfcal{E}_{\hat n}(\tilde s,\kappa,\mu)=\sum_{p=0}^\infty \mathbfcal{I}_{p,\hat n}(\tilde s,\kappa,\mu)y^{1-p-\tilde s}{\Bigg(\sum_{c=1}^\infty c^{-2(p+ \tilde s)}K(-\hat n,\mu;c)}\Bigg),\label{3.8u}
 \end{align}
 
\begin{table}[htb!]
\centering
\rowcolors{2}{gray!15}{white}
\begin{tabular}{|c|c|c|}
\hline
\textbf{Parameter(s)} & \textbf{Summand} & \textbf{Resulting Sum} \\
\hline
$j=0,\,J=0,\,s_1=0,\,w=1$ & $\sum_{s=1}^{\infty}\phi(s)\,s^{-4}$  & $\dfrac{\zeta(3)}{\zeta(4)}$ \\[6pt]
$j=0,\,J=0,\,s_1=1,\,w=1$ & $\sum_{s=1}^{\infty}\phi(s)\,s^{-6}$  & $\dfrac{\zeta(5)}{\zeta(6)}$ \\[6pt]
$j=0,\,J=0,\,s_1=2,\,w=0$ & $\sum_{s=1}^{\infty}\phi(s)\,s^{-6}$  & $\dfrac{\zeta(5)}{\zeta(6)}$ \\[6pt]
$j=0,\,J=0,\,s_1=2,\,w=1$ & $\sum_{s=1}^{\infty}\phi(s)\,s^{-8}$  & $\dfrac{\zeta(7)}{\zeta(8)}$ \\[6pt]
$j=0,\,J=0,\,s_1=2,\,w=2$ & $\sum_{s=1}^{\infty}\phi(s)\,s^{-10}$ & $\dfrac{\zeta(9)}{\zeta(10)}$ \\[6pt]
$j=0,\,J=1,\,s_1=0,\,w=1$ & $\sum_{s=1}^{\infty}\mu(s)\,s^{-4}$  & $\dfrac{1}{\zeta(4)}$ \\[6pt]
$j=0,\,J=1,\,s_1=1,\,w=1$ & $\sum_{s=1}^{\infty}\mu(s)\,s^{-6}$  & $\dfrac{1}{\zeta(6)}$ \\[6pt]
$j=0,\,J=1,\,s_1=2,\,w=0$ & $\sum_{s=1}^{\infty}\mu(s)\,s^{-6}$  & $\dfrac{1}{\zeta(6)}$ \\[6pt]
$j=0,\,J=1,\,s_1=2,\,w=1$ & $\sum_{s=1}^{\infty}\mu(s)\,s^{-8}$  & $\dfrac{1}{\zeta(8)}$ \\[6pt]
$j=0,\,J=1,\,s_1=2,\,w=2$ & $\sum_{s=1}^{\infty}\mu(s)\,s^{-10}$ & $\dfrac{1}{\zeta(10)}$ \\[6pt]
\hline
\end{tabular}
\caption{Modular sum for different choices of the parameters.}
\label{Tab3}
\end{table}

\noindent
For $p=0,\hat n=0$ we can write,
\begin{align}
   \mathbfcal{E}_{0}(\tilde s,\kappa,\mu)=
\begin{cases}
y^{1-\tilde s}\sqrt{\pi}\frac{\Gamma(\tilde s-1/2)}{\Gamma(\tilde s)}\frac{\zeta(-\tilde s-3)}{\zeta(-\tilde s-2)} ,\hspace{2.5 cm}\texttt{for $\mu=0$}\\
\sqrt{\pi}\frac{\Gamma(\tilde s-1/2)}{\Gamma(\tilde s)}y^{1-\tilde s}\frac{1}{\zeta(2\tilde s)}\hspace{4.1 cm}\texttt{for $\mu=\pm 1$}\\
\sqrt{\pi}\frac{\Gamma(\tilde s-1/2)}{\Gamma(\tilde s)}\times y^{1-\tilde s}\frac{1}{\zeta(2\tilde s)}\Bigg(1+2^{1-2\tilde s}\Bigg)\hspace{1.5 cm}\texttt{for $\mu=\pm 2$}
\end{cases}
\end{align}
Therefore,
\begin{align}
\begin{split}
    \mathbfcal{E} (\kappa,\mu,s,w)=\frac{\,E_2^s(\tau)E_2^w(\bar \tau)}{|\eta(\tau)|^2}&\times \Bigg(e^{2\pi(\kappa y+i\mu x)}+\mathbfcal{E}_{0}(1-s-w,\kappa,\mu)\\&\hspace{3 cm}
    +\texttt{finite $\hat n$ corrections}\Bigg)
    \end{split}
\end{align}
For $\hat n =0$  and $s_1\in 0,1,2 \text{\,\,and\,\,} w\in 0,1,2$ only the sum is finite while $s_1=0,w=0$. Otherwise it encounters a power-law divergence. A few values of the sum are casted in Table \eqref{Tab3}. While it is important to investigate what happens for large $\hat{n}$, we leave it as an important future direction.
\bibliography{ref3}
\bibliographystyle{utphysmodb}
\end{document}